\documentclass[12pt]{article}
\usepackage{amsbsy}
\usepackage{amsfonts}
\usepackage{amsmath}
\usepackage{amssymb}
\usepackage{amsthm}
\usepackage{mathrsfs}
\usepackage{apacite}
\usepackage{setspace}
\usepackage{graphicx}
\usepackage{xcolor}
\usepackage{bm}
\usepackage{booktabs}
\usepackage[english]{babel}
\usepackage{natbib}
\usepackage[T1]{fontenc}
\usepackage[utf8]{inputenc}
\usepackage{authblk}
\usepackage[margin=1.1in]{geometry}
\usepackage{threeparttable}

\newtheorem{proposition}{Proposition}
\newtheorem{definition}{Definition}

\newtheorem{lemma}{Lemma}

\newtheorem*{runningexample}{Running example}
\newtheorem*{runningexamplerev}{Running example}

\newcommand{\Y}{\textbf{Y}}
\newcommand{\bth}{\bm\theta}

\title{Prior Probabilities for Constrained Hypotheses: Invariance, Neutrality, and Model Search}

\author{Joris Mulder}
\affil{Department of Methodology and Statistics, Tilburg University}
\date{\today}

\begin{document}
\maketitle

\begin{abstract}%
When testing hypotheses with equality and inequality constraints using Bayes factors, the hypotheses
are commonly assigned equal prior probabilities. The resulting posterior probabilities depend on how
the analyst divides the parameter space into hypotheses: splitting one hypothesis into two changes
the posterior probabilities of the other hypotheses, and a hypothesis can receive more posterior
probability than a hypothesis that contains it. An alternative is considered where the prior
probabilities of the hypotheses follow from a single prior distribution on the parameters. Prior
probabilities of the equality constrained subspaces are taken from the Bayesian variable selection
literature, and prior probabilities of the inequality constrained regions within these subspaces
are given by their relative complexities under the encompassing prior. Under this complexity-induced
prior, posterior probabilities do not depend on how the hypotheses are divided, are coherent across
nested hypotheses, and agree with those from an exploratory model search over the same subspaces. In
return, inequality constraints do not increase the posterior probability of a hypothesis relative
to the null hypothesis. The prior is therefore suited for analysts who want to test constrained
hypotheses without favoring specific constraints a priori, while the potential gain in evidence for the constraints
remains quantified by the Bayes factor.
\end{abstract}

%

Researchers often have expectations about the direction or the ordering of effects, for instance
that one predictor has a larger positive effect than another predictor. Such expectations can be
formulated as hypotheses with equality and inequality constraints on the parameters of a statistical
model and tested using Bayes factors \citep{klugkist2005,hoijtink2012,mulder2021bfpack}, for example
with the R packages BFpack \citep{mulder2021bfpack} or bain \citep{gu2018,hoijtink2019}.
A typical analysis contains a small number of hypotheses based on competing theories, such as a null
hypothesis, one or more hypotheses with order constraints, and their complement. The results are
reported as Bayes factors and as posterior probabilities of the hypotheses, the latter based on equal prior probabilities by default.

The choice of the prior probabilities for hypotheses in these testing problems has received very
little attention in the literature, unlike priors for the parameters
\citep[e.g.,][]{jeffreys1961,bergerpericchi2001,liang2008,mulder2010,bayarri2012,consonni2018,mulder2023bayes}.
Prior probabilities of models have mainly been studied in the variable selection literature
\citep{scottberger2010,chipman2001,george2010}, and hardly for hypotheses with inequality
constraints \citep[except, for instance, by][]{klaassen2020}. An analyst could refrain
from specifying prior probabilities and only report Bayes factors. Bayes factors, however, compare
two hypotheses at a time, whereas posterior probabilities quantify the uncertainty about all
hypotheses jointly after observing the data. Posterior probabilities are also reported by default
in software \citep{mulder2021bfpack,gu2018,garcia2018bayesian,van2020tutorial}, and they are
directly affected by the prior probabilities. Equal prior probabilities are the common default,
presumably because the posterior odds then equal the Bayes factors and because the choice appears
neutral.

As an example, consider two standardized regression effects, $\beta_1$ and $\beta_2$, and an analyst
who tests the order hypothesis `$\beta_1>\beta_2>0$' against the null hypothesis `$\beta_1=\beta_2=0$'
and the complement, with default prior probabilities of $\frac{1}{3}$ each. A second analyst tests the same
null hypothesis against the two-sided alternative $(\beta_1,\beta_2)\neq(0,0)$, with default prior probabilities
set to $\frac{1}{2}$. In the first application considered in this paper, using the adjusted
fractional Bayes factor in BFpack, the posterior probability of the null hypothesis is .86 for the
first analyst and .97 for the second. Moreover, the first analyst obtains a posterior probability of
.12 for $\beta_1>\beta_2>0$, and the second analyst obtains .03 for $(\beta_1,\beta_2)\neq(0,0)$,
even though the first hypothesis implies the second. With the JZS Bayes factor
\citep{roudermorey2012}, the differences are larger: .47 and .81 for the null hypothesis, and .46
and .19 for the two alternatives.

Equal prior probabilities thus appear neutral, but the resulting posterior probabilities depend on
how the analyst divides the parameter space into hypotheses: the posterior probability of a null
hypothesis depends on whether it is tested against the complete two-sided parameter space or whether this two-sided parameter space is split between two hypotheses, such as an inequality constrained subspace and the complement. Moreover,
a hypothesis with inequality constraints can receive more posterior probability than
a hypothesis that contains it. \citet{stern2005}, in a discussion of \citet{klugkist2005}, showed
this dependence for an analysis of variance with four groups: the posterior probability of the
ordering $\mu_1<\mu_2<\mu_3<\mu_4$ decreased from .80 to .31 when more orderings were added to the
set of hypotheses. \citet{stern2005} proposed to use posterior probabilities on the parameter space
under the unconstrained model instead, which do not have these properties. This proposal has two limitations. First, it cannot be used for hypotheses with equality constraints \citep{klugkist2005reply}, which are common in practice. Second, these posterior
probabilities do not by themselves quantify the evidence in the data: a hypothesis that covers a
larger part of the parameter space, such as the complement of `$\beta_1>\beta_2>0$', receives more
posterior probability by construction. The evidence for an inequality constrained hypothesis is
quantified by the Bayes factor, which balances relative fit against relative complexity \citep{klugkist2005,mulder2014}.

In this paper the prior probabilities of the hypotheses are derived from a single prior
distribution on the parameters. This prior consists of two layers. The first layer assigns prior
probabilities to the equality constrained subspaces, following the priors used in Bayesian variable
selection \citep{scottberger2010}. The second layer assigns prior probabilities to the inequality
constrained regions within these subspaces, given by their relative complexities under the
encompassing prior \citep{mulder2014,klaassen2020}. I refer to the result as the complexity-induced
prior. For hypotheses with only inequality constraints, the resulting posterior probabilities are the
posterior probabilities on the parameter space of \citet{stern2005}. Unlike in \citet{stern2005},
however, these posterior probabilities do not replace the Bayes factors.

Using the complexity-induced prior, the posterior probabilities of the hypotheses do not depend on how the
hypotheses are divided, are coherent across nested hypotheses, and agree with the posterior
probabilities from an exploratory model search over the same subspaces. In return, inequality
constraints do not increase the posterior probability of a hypothesis relative to the null
hypothesis, while their support remains visible in the Bayes factor. The complexity-induced prior is
thus intended for analysts who want to be neutral about the constraints a priori: no prior belief
about, for instance, the direction of an effect is added beyond what is implied by the prior on the
parameters, and the gain in evidence from testing inequality constrained hypotheses is retained in
the Bayes factor.

The paper is organized as follows. First the formulation of the analyst's hypotheses is given, the parameter priors, and a
running example is introduced. Then the complexity-induced prior is introduced. Next, the posterior
probabilities, an empirical illustration, and three implications are discussed. The paper ends with recommendations for
practice and a discussion.

\section{Hypotheses, Parameter Priors, and Running Example}\label{sec:setup}

\subsection{The Analyst's Hypothesis Space}\label{sec:hypspace}

Let $\bth$ collect the $p$ parameters of interest, with unconstrained parameter space $\Theta_u$. On
this parameter vector, the analyst has formulated $T$ competing hypotheses with equality and/or
inequality (one-sided, order) constraints based on competing scientific theories or expectations.

\begin{definition}
The analyst's hypothesis space, denoted by $\mathscr{H}$, is a set of $T$ competing hypotheses of
the form
\begin{equation}\label{eq:hyp}
\mathcal{H}_{t}:\ \textbf{R}^E_{t}\bth=\textbf{r}^E_{t} \quad\&\quad
\textbf{R}^I_{t}\bth>\textbf{r}^I_{t},
\end{equation}
or unions of such hypotheses (e.g., the complement), where $[\textbf{R}^E_t|\textbf{r}^E_t]$ and
$[\textbf{R}^I_t|\textbf{r}^I_t]$ define the equality and inequality constraints of
$\mathcal{H}_t$, for $t=1,\ldots,T$. The hypotheses are assumed to be nonoverlapping, but they do
not need to cover the entire parameter space.
\end{definition}

The equality constraints of a hypothesis of the form \eqref{eq:hyp} define a linear subspace, which
will be referred to as its equality structure and index by $s$,
\begin{equation}\label{eq:hypE}
\mathcal{M}_s:\quad \textbf{R}^E_s\bth=\textbf{r}^E_s,
\end{equation}
with subspace $\Theta_s$ and free parameters $\bth_s$. Here, $\mathcal{M}$ is used to denote equality structures
to distinguish them from the analyst's hypotheses $\mathcal{H}_t$. The inequality constraints then
define a region $\Theta_{s,j}\subseteq\Theta_s$ within this equality structure, indexed by $j$. Moreover, $\mathcal{H}_{s,j}$ is used for the hypothesis that corresponds to the pair $(s,j)$. For an equality structure without inequality constraints, $j=0$, with
$\Theta_{s,0}=\Theta_s$. Each hypothesis in the analyst's hypothesis space is either of this form or a union of such hypotheses, such as the complement, written as
$\mathcal{H}_t=\cup_{(s,j)\in\mathcal{J}_t}\mathcal{H}_{s,j}$, where $\mathcal{J}_t$ contains the
pairs of equality structures and inequality regions that make up $\mathcal{H}_t$.

As the hypotheses are formulated based on competing theories, the default choice is to assign equal
prior probabilities,
\begin{equation}\label{eq:uniform}
P_{unif}(\mathcal{H}_t|\mathscr{H})=\frac{1}{T}.
\end{equation}
Even though this choice seems neutral, it depends on the (subjective) choice of the hypotheses by the analyst.

\subsection{Parameter Priors}\label{sec:parpriors}

A prior for the free parameters under each equality structure has been chosen,
denoted by $\pi_s(\bth_s)$. In the regression setting, for instance, common default choices include the $g$-prior
\citep{zellner1986}, mixtures of $g$-priors \citep{zellnersiow1980,liang2008,roudermorey2012}, and
fractional priors \citep{ohagan1995,mulder:2014b,mulderolsson2019}. The literature on such parameter
priors is extensive \citep[e.g.,][and the references therein]{jeffreys1961,kassraftery1995,bergerpericchi1996,bergerpericchi2001,bayarri2012,consonni2018}. These parameter priors determine the marginal likelihoods,
\[
m(\Y\,|\,\mathcal{M}_s)=\int p(\Y|\bth_s)\pi_s(\bth_s)\,d\bth_s,
\]
and thereby the Bayes factors between equality structures.

Under an inequality constrained region, the prior follows as a truncation of the prior under the
equality structure, which is known as the encompassing prior approach \citep{klugkist2005}. Quoting
Leucari and Consonni (2003) and Roverato and Consonni (2004): ``If nothing was elicited to indicate
that the two priors should be different, then it is sensible to specify [the prior of the inequality
constrained hypothesis] to be \ldots\ as close as possible to [the prior of the alternative
hypothesis]'' \citep{hoijtink2013}. 
Thus,
\[
\pi_{s,j}(\bth_s) = \pi_{s}(\bth_s) I(\bth_s\in \Theta_{s,j}) c_{s,j}^{-1},
\]
where the normalizing constant is the `relative complexity' of the inequality constrained hypothesis \citep{mulder2014},
\begin{equation}\label{eq:c}
c_{s,j}=\Pr(\bth_s\in\Theta_{s,j}\,|\,\mathcal{M}_{s}),
\end{equation}
i.e., the prior probability that the inequality constraints hold within equality structure $s$.
Similarly, the `relative fit'
\begin{equation}\label{eq:f}
f_{s,j}(\Y)=\Pr(\bth_s\in\Theta_{s,j}\,|\,\mathcal{M}_{s},\Y)
\end{equation}
is the posterior probability that they hold, with $c_{s,0}=f_{s,0}=1$. The Bayes factor of
$\mathcal{H}_{s,j}$ against its equality structure is then
\begin{equation}\label{eq:BF}
B_{(s,j),s}=\frac{m_{s,j}(\Y)}{m(\Y\,|\,\mathcal{M}_s)}=\frac{f_{s,j}(\Y)}{c_{s,j}},
\end{equation}
implying a maximal value of $c_{s,j}^{-1}$ \citep{mulder2010}.
Under equal prior probabilities \eqref{eq:uniform}, the posterior odds of two hypotheses equal their
Bayes factor.

\begin{runningexample}
Two applications are considered which have the same types of hypotheses on two regression effects $\beta_1$
and $\beta_2$, with standardized variables and while controlling for covariates.
In the first application, using the data of the first experiment of Carlsson and Sinclair (2017), 
the outcome is the perception that a female job applicant was
discriminated against, and $\beta_1$ and $\beta_2$ are the effects of feminist identification and
stigma consciousness, controlling for gender and the beliefs that women and that men are generally
discriminated against. Under the same-gender bias explanation, identification with the victim, as
operationalized by these two predictors, increases the perception of discrimination \citep{mulderolsson2019}.
In the second application, using the European Values Study data for Germany, the outcome is the attitude toward immigrants, and $\beta_1$ and $\beta_2$
are the effects of educational level and social class, controlling for income and gender. Ethnic
competition theory predicts a positive effect of both, and education may be the most important
factor because of its role in shaping one's identity \citep{mulderraftery2022}.
In both applications, the constrained hypotheses have the following forms
\begin{align*}
\mathcal{H}_{1}&:\beta_1=\beta_2=0,\\
\mathcal{H}_{2}&:\beta_1>\beta_2>0,\\
\mathcal{H}_{3}&:\text{neither }\mathcal{H}_1\text{ nor }\mathcal{H}_2
\end{align*}
Under the uniform prior in \eqref{eq:uniform}, each receives a prior probability of $\frac{1}{3}$. The first two consist
of a single pair $(s,j)$. $\mathcal{H}_1$ has the equality structure $\mathcal{M}_{00}$ with
$\Theta_{00}=\{(0,0)\}$, so that $\mathcal{J}_1=\{(00,0)\}$, and $\mathcal{H}_2$ lies in the equality
structure without equality constraints, $\mathcal{M}_{uu}$ with $\Theta_{uu}=\mathbb{R}^2$, so that
$\mathcal{J}_2=\{(uu,1)\}$. The complement contains the remaining region under $\mathcal{M}_{uu}$,
$\Theta_{uu,2}=\Theta_{uu}\setminus\Theta_{uu,1}$, and possibly other pairs $(s,j)$, which
depends on the equality structures that are considered plausible a priori (see the following subsection on equality
structures with nonzero probability). For a
parameter prior that is centered at zero with a covariance structure proportional to that of the
estimates, as for the AFBF \citep{mulder2022bayesian}, the relative complexity of $\mathcal{H}_2$ is $c_{uu,1}=.102$ in the first application
and $c_{uu,1}=.087$ in the second, so that the Bayes factor $B_{(uu,1),uu}=f_{uu,1}(\Y)/c_{uu,1}$
is at most about 10 and 11, respectively. The JZS Bayes factor \citep{roudermorey2012} gives
practically the same relative complexities.
\end{runningexample}

\section{Complexity-Induced Prior Probabilities}\label{sec:ciprior}

The hypotheses are specified on two levels, an equality constrained part and an inequality
constrained part on the remaining free parameters. This structure is used to construct their prior
probabilities: one layer for the equality structures, following the literature on Bayesian variable
selection, and one layer for the inequality constrained regions, following from the parameter priors.

\subsection{Equality Structures With Nonzero Probability}\label{sec:S}

A continuous prior on $\bth$, as used in estimation, assigns zero probability to
any specific parameter value, such as $\theta=0$, which is in line with the argument of
\citet{Meehl1967} that a parameter value is practically never exactly 0. On the other hand, there are three reasons to
assign a nonzero probability nevertheless. First, \citet{Johnson2017} estimated that more than 90\%
of null hypotheses tested in psychology may actually be true. Second, equality constraints give a
parsimonious solution: when there is no evidence that an effect exists, it may be useful to assume
it to be exactly zero. This is also done in Bayesian variable selection \citep{scottberger2010},
where an effect equals zero with nonzero probability, as in the spike-and-slab prior
\citep{ishwaran2005spike}, and in testing the equality of effects \citep{mulderraftery2022} or of
group means \citep{vandenbergh2026}. Third, when the analyst formulates a hypothesis with equality
constraints, this suggests that the constraints are plausible a priori; otherwise there would be no
reason for testing them.

To formalize this, I denote by $\mathcal{S}$ the set of equality structures that receive nonzero probability, with
$P(\mathcal{M}_s)>0$ and $\sum_{s\in\mathcal{S}}P(\mathcal{M}_s)=1$. The set $\mathcal{S}$
is determined by the equality constraints that the analyst deems plausible a priori, referred
to as the candidate equality constraints: $\mathcal{S}$ consists of the distinct subspaces that are
obtained by imposing any subset of the candidate constraints, including the empty set, which gives
$\Theta_u$. By default, the candidates are the separate equality constraints in the analyst's
hypotheses (e.g., $(\beta_1,\beta_2)=(0,0)$ gives the candidates $\beta_1=0$ and $\beta_2=0$), but the
analyst can add other candidates, such as $\beta_1=\beta_2$, when these are considered plausible as
well. Equality constrained subspaces that are not in $\mathcal{S}$ receive zero prior probability.

\subsection{The Equality Layer}\label{sec:eqlayer}

For each equality structure $\mathcal{M}_s$, let $r_s$ denote the number of parameters that are
removed by its equality constraints, i.e., the rank of $\textbf{R}^E_s$. Furthermore, let $R$ be the
number of parameters removed when all candidate equality constraints are imposed, and let $N_r$ be
the number of equality structures in $\mathcal{S}$ that remove $r$ parameters. The following sets the prior probability for $\mathcal{M}_s$,
\begin{equation}\label{eq:eqlayer}
P(\mathcal{M}_s)=\frac{P(r_s)}{N_{r_s}},
\end{equation}
i.e., a prior on the number of removed parameters which is equally divided over the equality
structures that remove this number of parameters. A binomial distribution with size $R$ and
probability $\delta$ for $P(r)$ gives
\begin{equation}\label{eq:eqlayerbin}
P(\mathcal{M}_s)=\binom{R}{r_s}\frac{\delta^{r_s}(1-\delta)^{R-r_s}}{N_{r_s}}.
\end{equation}
For a small number of hypotheses, $\delta$ is typically fixed (e.g., at $\frac{1}{2}$), which
provides no multiplicity control. The correction is to place a beta prior on $\delta$, resulting in a
beta-binomial distribution for $P(r)$ \citep{scottberger2010}. For a uniform prior on $\delta$,
$P(\mathcal{M}_s)=1/\{(R+1)N_{r_s}\}$. An interval hypothesis that represents a negligible effect, such as $|\theta|<a$, can be treated as
an equality constraint: its prior probability then follows from $\delta$, which reflects the prior
probability that an effect is practically negligible.

Counting removed parameters, rather than assigning a probability to each candidate constraint, is
needed when the candidate constraints are dependent. For example, for the candidate constraints
$\beta_1=0$, $\beta_2=0$, and $\beta_1=\beta_2$, any two constraints imply the third, so that exactly
two constraints cannot hold. The equality structures are then the plane ($r=0$), the three lines
($r=1$), and the origin ($r=2$), which receive prior probabilities of $\frac{1}{4}$, $\frac{1}{6}$
each, and $\frac{1}{4}$ for $\delta=\frac{1}{2}$. Other linear equality constraints, such as
$\beta_2-\beta_1=\beta_3-\beta_2$ for a linear change over time \citep{mulder2009bayesian}, are included as additional candidate
constraints in the same way. Expression \eqref{eq:eqlayer} should be seen as a default choice, which
can be replaced by another prior on $\mathcal{S}$; the results below do not depend on this choice.

\subsection{The Complexity-Induced Prior}\label{sec:prior}

Together with the parameter priors $\pi_s$, the prior probabilities of the equality structures yield
a complete prior distribution on $\bth$: first an equality structure $s$ is drawn with probability
$P(\mathcal{M}_s)$, and subsequently the free parameters are drawn from $\pi_s$. Under this prior,
the probability of a hypothesis $\mathcal{H}_{s,j}$ is $P(\mathcal{M}_s)\,c_{s,j}$, and
the probability of a hypothesis in the analyst's hypothesis space is
\begin{equation}\label{eq:Pi}
\Pi_t=\sum_{(s,j)\in\mathcal{J}_t}P(\mathcal{M}_s)\,c_{s,j}.
\end{equation}

\begin{definition}
The complexity-induced prior probability of a hypothesis in the analyst's hypothesis space is
\begin{equation}\label{eq:cpp}
P_{comp}(\mathcal{H}_t|\mathscr{H})=\frac{\Pi_t}{\sum_{t'}\Pi_{t'}}.
\end{equation}
\end{definition}

The hypotheses are exhaustive when $\sum_t\Pi_t=1$, i.e., when for each equality structure in
$\mathcal{S}$ the relative complexities of the regions included in the hypotheses sum to one. In that
case $P_{comp}(\mathcal{H}_t|\mathscr{H})=\Pi_t$. Otherwise, the remaining prior probability belongs
to parameter values that the analyst excluded, and \eqref{eq:cpp} conditions on the analyst's
hypothesis space. For hypotheses with only inequality constraints, \eqref{eq:cpp} reduces to the
relative complexities, as discussed by \citet{mulder2014} and \citet{klaassen2020}.

The idea is thus to use one coherent prior: the prior probabilities of the inequality parts of the hypotheses are not chosen
manually but follow from the prior on the parameters. A related observation was made by
\citet{mulder2014} for the parameter prior: when the prior is constructed such that all orderings
of the parameters are equally likely a priori, the Bayes factor depends on the parameterization of
the model, which is avoided when the prior has the covariance structure of the data. Likewise, the
prior probabilities of the hypotheses should follow directly from the encompassing prior, rather
than be specified manually (such as setting equal prior probabilities) to ensure consistency.

\begin{runningexamplerev}
As the two effects $\beta_1$ and $\beta_2$ are fixed to zero under $\mathcal{H}_1$, the set of equality structures
$\mathcal{S}$ contains
$\mathcal{M}_{uu}$, $\mathcal{M}_{0u}$ with $\Theta_{0u}=\{(0,\beta_2)\}$, $\mathcal{M}_{u0}$ with
$\Theta_{u0}=\{(\beta_1,0)\}$, and $\mathcal{M}_{00}$. The complement therefore consists of the
remaining region under $\mathcal{M}_{uu}$ and the two equality structures where one effect is zero,
$\mathcal{J}_3=\{(uu,2),(0u,0),(u0,0)\}$. With $R=2$, $N_0=N_2=1$, and $N_1=2$, the binomial prior
with $\delta=\frac{1}{2}$ gives $P(\mathcal{M}_s)=\frac{1}{4}$ for all four equality structures, so
that
\begin{align*}
\Pi_1&=P(\mathcal{M}_{00})=.250,\\
\Pi_2&=P(\mathcal{M}_{uu})\,c_{uu,1}=.026,\\
\Pi_3&=P(\mathcal{M}_{uu})\,c_{uu,2}+P(\mathcal{M}_{0u})+P(\mathcal{M}_{u0})=.724,
\end{align*}
in the first application, and $.250$, $.022$, and $.728$ in the second. These sum to one, as the
hypotheses are exhaustive. Compared to the uniform prior, the complement receives a considerably
larger prior probability.
\end{runningexamplerev}

The complement is defined here with respect to the entire parameter space, and thus it also contains
the equality structures $\mathcal{M}_{0u}$ and $\mathcal{M}_{u0}$, which do not appear in the
analyst's hypotheses. Alternatively, the complement could be defined under $\mathcal{M}_{uu}$ only,
$\mathcal{J}_3=\{(uu,2)\}$. This is how the complement is defined in BFpack
\citep{mulder2021bfpack}, where the hypotheses then receive equal prior probabilities by default.
Under the complexity-induced prior, the hypotheses would not be exhaustive with this complement,
and \eqref{eq:cpp} would give prior probabilities of $.500$, $.051$, and $.449$ in the first
application. Throughout this paper the first definition is used, as it is consistent with the
equality structures in $\mathcal{S}$ having nonzero probability. Moreover, if the analyst adds a
hypothesis on the excluded equality structures afterwards (e.g., $\beta_1=0$), this only splits the
prior probability of the complement and leaves the prior probabilities of $\mathcal{H}_1$ and
$\mathcal{H}_2$ unchanged (see Implication~I).

\section{Posterior Probabilities and Implications}\label{sec:impl}

The posterior probability of equality structure $s$ is
\[
P(\mathcal{M}_s|\Y)=\frac{P(\mathcal{M}_s)\,m(\Y\,|\,\mathcal{M}_s)}{\sum_{s'\in\mathcal{S}}P(\mathcal{M}_{s'})\,m(\Y\,|\,\mathcal{M}_{s'})},
\]
and analogous to \eqref{eq:Pi}, it follows that
\begin{equation}\label{eq:Pipost}
\Pi_t(\Y)=\sum_{(s,j)\in\mathcal{J}_t}P(\mathcal{M}_s|\Y)\,f_{s,j}(\Y)
\end{equation}
for the posterior probability that the constraints of $\mathcal{H}_t$ hold. The prior under
$\mathcal{H}_t$ is the complete prior distribution restricted to $\mathcal{H}_t$, i.e., a mixture of
the truncated priors $\pi_{s,j}$, $(s,j)\in\mathcal{J}_t$, with weights
$P(\mathcal{M}_s)c_{s,j}/\Pi_t$. Under the uniform prior, the complement is defined under the
unconstrained equality structure $\mathcal{M}_u$, as is common in practice, so that each hypothesis consists of a single pair $(s,j)$.

\begin{lemma}\label{lem:post}
Under the complexity-induced prior \eqref{eq:cpp},
\begin{equation}\label{eq:postcomp}
P_{comp}(\mathcal{H}_{t}|\Y,\mathscr{H})=\frac{\Pi_t(\Y)}{\sum_{t'}\Pi_{t'}(\Y)},
\end{equation}
which equals $\Pi_t(\Y)$ when the hypotheses are exhaustive, and in particular for a hypothesis $\mathcal{H}_{s,j}$,
\begin{equation}\label{eq:fact2}
P_{comp}(\mathcal{H}_{s,j}|\Y,\mathscr{H})=P(\mathcal{M}_s|\Y)\,f_{s,j}(\Y).
\end{equation}
Under the uniform prior \eqref{eq:uniform},
\begin{equation}\label{eq:postunif}
P_{unif}(\mathcal{H}_{s,j}|\Y,\mathscr{H})=\frac{B_{(s,j),u}}{\sum_{(s',j')}B_{(s',j'),u}},
\quad\text{with}\quad
B_{(s,j),u}=\frac{m(\Y\,|\,\mathcal{M}_s)}{m(\Y\,|\,\mathcal{M}_{u})}\,\frac{f_{s,j}(\Y)}{c_{s,j}},
\end{equation}
where the sum is over the hypotheses in $\mathscr{H}$.
\end{lemma}

Under the complexity-induced prior, the posterior probability of a hypothesis is thus the posterior
probability that its constraints hold, and the relative complexities do not appear. Under the uniform
prior, the posterior probabilities are normalized Bayes factors. The following result compares
posterior probabilities across hypothesis spaces. The proofs of Lemma~\ref{lem:post} and
Proposition~\ref{prop:inv} are given in Appendix~A.

\begin{proposition}\label{prop:inv}
Let $\mathscr{H}$ and $\mathscr{H}^*$ be two hypothesis spaces based on the same set of equality
structures $\mathcal{S}$, with the same prior probabilities $P(\mathcal{M}_s)$ and parameter priors
$\pi_s$, such that for each $s\in\mathcal{S}$ the regions included in the hypotheses of $\mathscr{H}$
and $\mathscr{H}^*$ have the same union. Under the complexity-induced prior,
\begin{enumerate}
\item[(i)] a hypothesis that is contained in both $\mathscr{H}$ and $\mathscr{H}^*$ has the same
posterior probability;
\item[(ii)] when a hypothesis in $\mathscr{H}$ is the union of hypotheses in $\mathscr{H}^*$, its
posterior probability is the sum of their posterior probabilities;
\item[(iii)] when $\mathcal{H}_t\in\mathscr{H}$ and $\mathcal{H}_{t'}\in\mathscr{H}^*$ with
$\mathcal{J}_t\subseteq\mathcal{J}_{t'}$, then
$P_{comp}(\mathcal{H}_t|\Y,\mathscr{H})\leq P_{comp}(\mathcal{H}_{t'}|\Y,\mathscr{H}^*)$.
\end{enumerate}
\end{proposition}

In the proposition, the condition means that both hypothesis spaces cover the same part of the parameter space and only
divide it differently. Parts (i), (iii), and (ii) are discussed in Implications I, II, and III,
respectively, after the empirical illustration.

\begin{runningexamplerev}
Under the complexity-induced prior,
\begin{align*}
P_{comp}(\mathcal{H}_1|\Y,\mathscr{H})&=P(\mathcal{M}_{00}|\Y),\\
P_{comp}(\mathcal{H}_2|\Y,\mathscr{H})&=P(\mathcal{M}_{uu}|\Y)\,f_{uu,1}(\Y),\\
P_{comp}(\mathcal{H}_3|\Y,\mathscr{H})&=P(\mathcal{M}_{uu}|\Y)\,f_{uu,2}(\Y)+P(\mathcal{M}_{0u}|\Y)+P(\mathcal{M}_{u0}|\Y),
\end{align*}
while under the uniform prior the posterior probabilities are proportional to
$B_{1,uu}=m(\Y\,|\,\mathcal{M}_{00})/m(\Y\,|\,\mathcal{M}_{uu})$, $B_{2,uu}=f_{uu,1}(\Y)/c_{uu,1}$,
and $B_{3,uu}=f_{uu,2}(\Y)/c_{uu,2}$, where $\mathcal{M}_{0u}$ and $\mathcal{M}_{u0}$ play no role.
\end{runningexamplerev}

\subsection{Empirical Illustration}\label{sec:illus}

For the two applications, the posterior probabilities under the complexity-induced prior
and the uniform prior are compared. To illustrate the role of the division of the alternative, a
second analyst is considered who tests the same null hypothesis against the two-sided alternative,
$\mathscr{H}^*=\{\mathcal{H}^*_1,\mathcal{H}^*_2\}$ with $\mathcal{H}^*_1:\beta_1=\beta_2=0$ (which corresponds to $\mathcal{H}_1$ of the first analyst) and
$\mathcal{H}^*_2:(\beta_1,\beta_2)\neq(0,0)$. The candidate equality constraints, and thus
$\mathcal{S}$, are unchanged, and $\mathcal{H}^*_2$ is the union of $\mathcal{H}_2$ and
$\mathcal{H}_3$.

In the first application ($n=152$), the estimates are $\hat\beta_1=0.12$ ($SE=0.08$) and
$\hat\beta_2=0.06$ ($SE=0.09$), and in the second application ($n=1605$), $\hat\beta_1=0.16$
($SE=0.03$) and $\hat\beta_2=0.11$ ($SE=0.03$). The results are given for the adjusted fractional
Bayes factor (AFBF) in BFpack \citep{mulder2021bfpack} and for the JZS Bayes factor in BayesFactor
\citep{roudermorey2012}, with $P(\mathcal{M}_s)=\frac{1}{4}$. For both Bayes factors, the parameter prior is
centered at zero with a similar covariance structure, so that
the relative complexities are practically the same. 
The R code is given in Appendix~B.

\begin{table}
\caption{Posterior Probabilities Under the Complexity-Induced Prior and the Uniform Prior for the Two Applications}
\label{tab:inv}
\begin{threeparttable}
\small\setlength{\tabcolsep}{3pt}
\begin{tabular}{llcccccccccc}
\toprule
 & & \multicolumn{5}{c}{Complexity-induced} & \multicolumn{5}{c}{Uniform}\\
\cmidrule(lr){3-7}\cmidrule(lr){8-12}
 & & \multicolumn{3}{c}{$\mathscr{H}$} & \multicolumn{2}{c}{$\mathscr{H}^*$} & \multicolumn{3}{c}{$\mathscr{H}$} & \multicolumn{2}{c}{$\mathscr{H}^*$}\\
\cmidrule(lr){3-5}\cmidrule(lr){6-7}\cmidrule(lr){8-10}\cmidrule(lr){11-12}
Application & Method & $\mathcal{H}_1$ & $\mathcal{H}_2$ & $\mathcal{H}_3$ & $\mathcal{H}^*_1$ & $\mathcal{H}^*_2$ & $\mathcal{H}_1$ & $\mathcal{H}_2$ & $\mathcal{H}_3$ & $\mathcal{H}^*_1$ & $\mathcal{H}^*_2$\\
\midrule
Carlsson and Sinclair & AFBF & \textbf{.621} & .009 & .370 & \textbf{.621} & .379 & \textbf{.858} & .124 & .019 & \textbf{.967} & .033\\
 & JZS & \textbf{.429} & .042 & .528 & \textbf{.429} & .571 & \textbf{.471} & .457 & .072 & \textbf{.809} & .191\\
EVS & AFBF & \textbf{.000} & .820 & .180 & \textbf{.000} & 1.000 & \textbf{.000} & .986 & .014 & \textbf{.000} & 1.000\\
 & JZS & \textbf{.000} & .855 & .145 & \textbf{.000} & 1.000 & \textbf{.000} & .985 & .015 & \textbf{.000} & 1.000\\
\bottomrule
\end{tabular}
\begin{tablenotes}[para,flushleft]
{\small \textit{Note.} AFBF = adjusted fractional Bayes factor; JZS = JZS Bayes factor; EVS = European
Values Study. Estimates (with standard errors): $\hat\beta_1=0.12$ $(0.08)$ and $\hat\beta_2=0.06$
$(0.09)$ for Carlsson and Sinclair ($n=152$); $\hat\beta_1=0.16$ $(0.03)$ and $\hat\beta_2=0.11$
$(0.03)$ for EVS ($n=1605$). The null hypothesis is in bold. $\mathcal{H}^*_1=\mathcal{H}_1$ and
$\mathcal{H}^*_2=\mathcal{H}_2\cup\mathcal{H}_3$.}
\end{tablenotes}
\end{threeparttable}
\end{table}

In the first application both effects are small relative to their standard errors, and the null
hypothesis receives substantial posterior probability under both priors. The priors differ in how
the remaining posterior probability is divided. Under the complexity-induced prior it mostly goes to
the complement, which has a larger prior probability than the order hypothesis (.724 against .026)
and which contains the equality structures where only one of the effects is zero. Under the uniform
prior, the order hypothesis and the complement have the same prior probability and these equality
structures are not considered, so that the order hypothesis receives more posterior probability than
the complement. In the second application both effects are clearly present,
and $\mathcal{H}_2$ is supported under both priors. Note that the differences between the AFBF and the JZS
Bayes factor can be explained by the difference in the scale of the parameter prior.

\subsection{Implication I: Invariance Under Refinement}\label{sec:inv}

By Proposition~\ref{prop:inv}(i), the way in which the analyst divides the alternative into
inequality constrained hypotheses does not affect the posterior probabilities of the other
hypotheses. In the first application, the posterior probability of the null hypothesis is .62 (AFBF)
and .43 (JZS), whether it is tested against $\mathcal{H}_2$ and $\mathcal{H}_3$ or against the
two-sided alternative (Table~\ref{tab:inv}). Under the uniform prior, it is .86 and .97 using the
AFBF, and .47 and .81 using the JZS Bayes factor. Moreover, under the uniform prior the prior probability of the null
hypothesis, $1/T$, decreases with every hypothesis that is added to the alternative, whereas under
the complexity-induced prior it remains $P(\mathcal{M}_{00})$, so that dividing the alternative into
many hypotheses does not require a correction for multiplicity. \citet{stern2005} showed the same
dependence for an analysis of variance with four groups.

A consequence is that the alternative may be divided into inequality constrained hypotheses after
observing the data without affecting the posterior probability of the null hypothesis. Hence, a
decision rule based on the posterior probability of the null hypothesis has the same frequentist error
rates regardless of whether the division was chosen before or after observing the data. This contrasts
with the common notion that an analyst may not switch from a two-sided to a one-sided alternative
after observing the direction of the effect. It only concerns conclusions about the null hypothesis:
choosing the division after observing the data does affect the error rates of conclusions among the
alternatives. This also connects to e-values \citep{grunwald2024}: the Bayes factor of the union of
the alternatives against the null hypothesis does not depend on the division, so when it is an
e-variable, as for a simple null hypothesis, it remains one after refinement.

The invariance requires the same equality structures: adding a hypothesis with a new equality
constraint, such as $\beta_1=\beta_2>0$, changes $\mathcal{S}$ and therefore all posterior
probabilities. It also requires that the prior probabilities of refined hypotheses are proportional to
their relative complexities. Dividing the prior probability of the alternative equally, e.g.,
$P(\mathcal{H}_0)=\frac12$ and $\frac14$ for $\theta<0$ and $\theta>0$, is only consistent with
testing $\theta=0$ against $\theta\neq0$ when the relative complexities are equal. This does not
hold, for instance, for the (unadjusted) fractional Bayes factor \citep{ohagan1995}, where the prior is centered at the
MLE, or when testing multiple parameters that are correlated a priori such as the $g$-prior.

\subsection{Implication II: Coherence, and What Directional Testing Does (Not) Buy}\label{sec:coh}

By Proposition~\ref{prop:inv}(iii), a hypothesis cannot receive more posterior probability than a
hypothesis that contains it. In particular, for exhaustive hypotheses,
\begin{equation}\label{eq:nogain}
P_{comp}(\mathcal{H}_{s,j}|\Y,\mathscr{H})=P(\mathcal{M}_s|\Y)\,f_{s,j}(\Y)\leq P(\mathcal{M}_s|\Y).
\end{equation}
Under the uniform prior this does not hold. In the first application, both $\mathscr{H}$ and
$\mathscr{H}^*$ cover $\mathcal{M}_{uu}$ and $\mathcal{M}_{00}$ under the uniform approach, so that
both condition on the same event. Yet the uniform prior gives .12 to `$\beta_1>\beta_2>0$' and .03 to
`$(\beta_1,\beta_2)\neq(0,0)$' using the AFBF (.46 and .19 using the JZS Bayes factor), even though
the first implies the second, while the complexity-induced prior gives .01 and .38 (.04 and .57).
This is analogous to the conjunction
fallacy \citep{tverskykahneman1983}, where a conjunction (``Linda is a bank teller and active in the
feminist movement'') is judged more probable than one of its constituents (``Linda is a bank
teller''), which has been explained by judging how well a statement is supported by the evidence
rather than how probable it is \citep{crupi2008,tentori2013}. Under the uniform prior, the posterior
odds equal the Bayes factors, which quantify support, and a more specific hypothesis can be better
supported than a hypothesis that contains it \citep{lavineschervish1999}. Posterior probabilities under
the complexity-induced prior are coherent in the sense of \citet{gabriel1969}, see also
\citet{izbicki2015} and \citet{fossaluza2017}.

The price of this coherence is that inequality constraints do not increase the posterior probability
of a hypothesis relative to the null hypothesis. The Bayes factor $B_{(s,j),s}=f_{s,j}/c_{s,j}$ can
be as large as $1/c_{s,j}$ \citep{mulder2010}, e.g., 2 for a one-sided hypothesis or $p!$ for an ordering of $p$
exchangeable parameters, but the prior probability of $\mathcal{H}_{s,j}$ is smaller by the same
factor, so that the gain cancels in the posterior probabilities. By Proposition~\ref{prop:inv}(i),
directional testing therefore does not increase the power to reject the null hypothesis; the
inequality constraints provide information on where the parameters lie within an equality structure,
through $f_{s,j}$. This property was raised by \citet{klugkist2005reply} as an objection against the
posterior probabilities on the parameter space of \citet{stern2005}, and by \citet{mulder2014} as a
reason to recommend equal prior probabilities. Under the posterior probabilities of
\citet{stern2005}, the unconstrained model always has posterior probability one and a specific
ordering of many means remains improbable even when it is strongly supported by the Bayes factor.
Using the complexity-induced prior, the support for the constraints remains available through the Bayes factors, while the
posterior probabilities remain coherent, avoiding the conjunction fallacy.

Under the uniform prior the gain does reach the posterior probabilities, but only because more prior
probability is assigned to the directional hypothesis than its relative complexity. When testing
$\theta=0$, $\theta>0$, and $\theta<0$, the uniform prior assigns $\frac{1}{3}$ to $\theta>0$,
whereas the complexity-induced prior assigns $\frac14$ (for $\delta=\frac12$). The uniform prior thus
contains a prior belief about the direction that the analyst has not explicitly stated.

\subsection{Implication III: Confirmatory Testing as a Coarsening of Exploratory Search}\label{sec:collapse}

The literature on testing constrained hypotheses has developed largely separately from the literature
on Bayesian model selection. The first can be viewed as confirmatory, where a small set of hypotheses
of substantive interest is tested. The second can be viewed as exploratory, where the model space
consists of all models that can be constructed for the parameters, as in variable selection
\citep{georgemcculloch1993,liang2008,scottberger2010}, the selection of equal group means
\citep{vandenbergh2026}, or edge selection in graphical models \citep{mohammadi2015,williamsmulder2020,marsman2022}. \citet{stern2005} suggested to
consider all orderings of a set of means and to sum posterior probabilities over orderings, while
\citet{klugkist2005reply} argued that only theory-based hypotheses should be evaluated, to avoid models
without substantive meaning and capitalization on chance.

Consider the model space containing all equality structures in $\mathcal{S}$, each divided into the
regions defined by the inequality constraints in the analyst's hypotheses. Every hypothesis in
$\mathscr{H}$ is a union of models in this space, so that by Proposition~\ref{prop:inv}(ii) its
posterior probability under the complexity-induced prior equals the sum of the posterior probabilities
of these models. The confirmatory test and the exploratory search thus give consistent results. In
practice, existing model search algorithms can be used to obtain $P(\mathcal{M}_s|\Y)$, after which
the posterior probabilities of the hypotheses of interest follow from \eqref{eq:fact2}, with the
relative fits computed from posterior draws under each equality structure, and the complement
receiving the remaining posterior probability. The search therefore only needs to be performed over
the equality structures, e.g., $2^K$ inclusion patterns in variable selection with $K$ predictors
instead of $3^K$ signed models \citep{gu2022}, or the Bell number of partitions of $K$ group means
instead of the ordered Bell number (52 instead of 541 for $K=5$).

In variable selection, the candidate equality constraints are $\beta_k=0$, $k=1,\ldots,K$, so that
$R=K$, $N_r=\binom{K}{r}$, and \eqref{eq:eqlayerbin} becomes
$P(\mathcal{M}_s)=\delta^{r_s}(1-\delta)^{K-r_s}$, while a uniform prior on $\delta$ gives the prior
of \citet{scottberger2010}. For equality of means, the candidate constraints are $\theta_k=\theta_l$
for all pairs of $K$ means, the equality structures are the partitions of the means, $N_r$ is a
Stirling number of the second kind, and \eqref{eq:eqlayer} corresponds to the beta-binomial partition
prior of \citet{vandenbergh2026}. The one-sided variable selection of \citet{gu2022} with equal prior
probabilities for positive and negative effects coincides with the complexity-induced prior when each
sign pattern of $m$ included predictors has prior probability $2^{-m}$, e.g., for orthogonal
predictors, but not when the predictors are correlated.

The link with variable selection also implies that the multiplicity correction of
\citet{scottberger2010} carries over to the analyst's hypotheses: when a prior is placed on $\delta$,
the prior probabilities of the equality structures, and thereby of the hypotheses, are corrected for
the number of candidate equality constraints.

The equivalence requires the same equality structures and prior probabilities. A variable selection
over all predictors also considers the exclusion of covariates that are included in all of the
analyst's hypotheses, which can be avoided by including the covariates in all models. Finally, the
concern of \citet{klugkist2005reply} about capitalization on chance is partly addressed: the posterior
probabilities of the analyst's hypotheses are not affected by also exploring the model space, but
hypotheses that are formulated after inspecting the exploratory results should still be evaluated with
new data.

\section{Recommendations for Practice}\label{sec:rec}

\subsection{When to Use the Complexity-Induced Prior}

The complexity-induced prior is useful when the analyst has formulated a set of hypotheses with
equality and inequality constraints based on substantive or theoretical arguments, has chosen the parameter priors, but does not want to add
additional prior beliefs about the hypotheses: using the complexity-induced prior, the prior probabilities of the hypotheses follow naturally from these
choices. This results in neutral prior probabilities where a hypothesis that covers a smaller part of the
parameter space receives a smaller prior probability. The gain in evidence in favor of inequality constrained hypotheses that are supported by the data remains as Bayes factors are independent of prior beliefs. When the analyst has prior beliefs about the hypotheses, e.g., about the direction of one or more
effects (e.g., based on contextual knowledge or earlier published findings), prior probabilities can be specified accordingly. The complexity-induced prior would still be useful as a prior sensitivity check to assess the robustness of the findings.

\subsection{What to Report}

The posterior probabilities should be reported together with the prior probabilities and the Bayes factors. The posterior
probabilities give the probability that the constraints of each hypothesis hold, and the Bayes
factors show the gain in evidence from the constrained hypotheses, which would not be visible in the
posterior probabilities when using the complexity-induced prior. As the complexity-induced prior depends on the equality constraints that are deemed plausible a priori, it is recommended to report these as well.

\subsection{Choosing the Equality Structures and $\delta$}

By default, the candidate equality constraints are the separate equality constraints in the
analyst's hypotheses. Other candidates, such as $\beta_1=\beta_2$, are added when these are
considered plausible a priori as well. Covariates that are only controlled for, and on which no constraints are formulated, are kept in
the model under all equality structures and naturally no equality constraints would be considered on their effects. For a small number of hypotheses, $\delta$ can
be fixed, e.g., at $\frac{1}{2}$, which provides no multiplicity control. A uniform prior on $\delta$,
which gives $P(\mathcal{M}_s)=1/\{(R+1)N_{r_s}\}$, corrects for the number of candidate equality
constraints.

\subsection{Computation in BFpack and bain}

The posterior probabilities that BFpack \citep{mulder2021bfpack} and bain \citep{gu2019bain, hoijtink2019tutorial} report by default are based on equal prior probabilities,
with the complement defined under the unconstrained equality structure. The posterior probabilities
under the complexity-induced prior can be computed from the BFpack output in three steps (see
Appendix~B for the code of the two applications). First, the function \texttt{BF()} is applied to
the fitted model with the analyst's hypotheses, e.g.,
\texttt{hypothesis = "x1 > x2 > 0; x1 = x2 = 0"}, which gives the relative complexity and the relative
fit of the inequality constraints and the Bayes factor of the equality structure of the null
hypothesis against the unconstrained equality structure. Second, \texttt{BF()} is applied to each
remaining equality structure in $\mathcal{S}$, e.g., \texttt{hypothesis = "x1 = 0"}, which gives its
Bayes factor against the unconstrained equality structure. Third, these Bayes factors are combined
with the prior probabilities $P(\mathcal{M}_s)$ to obtain $P(\mathcal{M}_s|\Y)$, the posterior
probabilities of the hypotheses $\mathcal{H}_{s,j}$ follow from \eqref{eq:fact2}, and the complement receives
the remaining posterior probability. For analysts using bain, the computation is similar.

\section{Discussion}\label{sec:disc}

Equal prior probabilities for the hypotheses in the analyst's hypothesis space make the posterior
probabilities depend on how the parameter space is divided. The complexity-induced prior instead
derives the prior probabilities of the hypotheses from a single prior distribution on the parameters,
with prior probabilities of the equality structures from the variable selection literature and
relative complexities for the inequality constrained regions. The resulting posterior probabilities
are invariant to refinement, coherent across nested hypotheses, and consistent with exploratory model
search. At the same time, inequality constraints do not increase posterior probabilities relative to the null
hypothesis but their support is still reflected in the Bayes factors, which therefore remain important to
report.

The complexity-induced prior is intended for analysts who want to test constrained hypotheses using the Bayes factor while
being neutral about the plausibility of the constraints a priori. The uniform prior would not be the neutral choice: it
assigns the same prior probability to hypotheses that differ in size under the prior on the
parameters, which corresponds to a prior belief in favor of the more specific hypotheses. An analyst who wishes to translate their prior beliefs about the hypotheses, e.g., about the direction of an effect, can specify the prior probabilities accordingly, and the results under different prior choices can be considered next to each other.

The presented results also apply to testing problems beyond regression models, such as analysis of variance
\citep{klugkist2005}, contingency tables \citep{Klugkist:2010}, structural equation
models \citep{vanlissa2021}, network autocorrelation models \citep{dittrich2020network}, or multinomial models
\citep{heck2019,sarafoglou2023}: the construction only requires parameter priors under the equality
structures.
Because the posterior probabilities of the analyst's hypotheses follow from those of the equality
structures, output from existing model search algorithms can be used directly, so that confirmatory
and exploratory analyses of the same data can be reported side by side.

\subsection*{Transparency and Openness}
This study was not preregistered. No new data were collected for this study. Claude \citep{anthropic2026} was used to assist with editing parts of the text; the author takes full responsibility.

\subsection*{Acknowledgements}
The author was supported by an ERC Consolidator Grant (101087383).

\bibliographystyle{apacite}
\bibliography{refs_mulder_PM}

\begin{thebibliography}{}

\bibitem [\protect \citeauthoryear {%
{Anthropic}%
}{%
{Anthropic}%
}{%
{\protect \APACyear {2026}}%
}]{%
anthropic2026}
\APACinsertmetastar {%
anthropic2026}%
\begin{APACrefauthors}%
{Anthropic}.%
\end{APACrefauthors}%
\unskip\
\newblock
\APACrefYearMonthDay{2026}{}{}.
\newblock
\APACrefbtitle {Claude.} {Claude.}
\newblock
\APAChowpublished {Large language model}.
\newblock
\begin{APACrefURL} \url{https://claude.ai} \end{APACrefURL}
\PrintBackRefs{\CurrentBib}

\bibitem [\protect \citeauthoryear {%
Bayarri%
, Berger%
, Forte%
\BCBL {}\ \BBA {} Garc{\'\i}a-Donato%
}{%
Bayarri%
\ \protect \BOthers {.}}{%
{\protect \APACyear {2012}}%
}]{%
bayarri2012}
\APACinsertmetastar {%
bayarri2012}%
\begin{APACrefauthors}%
Bayarri, M\BPBI J.%
, Berger, J\BPBI O.%
, Forte, A.%
\BCBL {}\ \BBA {} Garc{\'\i}a-Donato, G.%
\end{APACrefauthors}%
\unskip\
\newblock
\APACrefYearMonthDay{2012}{}{}.
\newblock
{\BBOQ}\APACrefatitle {Criteria for {B}ayesian model choice with application to
  variable selection} {Criteria for {B}ayesian model choice with application to
  variable selection}.{\BBCQ}
\newblock
\APACjournalVolNumPages{The Annals of Statistics}{40}{3}{1550--1577}.
\newblock
\begin{APACrefDOI} \doi{10.1214/12-AOS1013} \end{APACrefDOI}
\PrintBackRefs{\CurrentBib}

\bibitem [\protect \citeauthoryear {%
Berger%
\ \BBA {} Pericchi%
}{%
Berger%
\ \BBA {} Pericchi%
}{%
{\protect \APACyear {1996}}%
}]{%
bergerpericchi1996}
\APACinsertmetastar {%
bergerpericchi1996}%
\begin{APACrefauthors}%
Berger, J\BPBI O.%
\BCBT {}\ \BBA {} Pericchi, L\BPBI R.%
\end{APACrefauthors}%
\unskip\
\newblock
\APACrefYearMonthDay{1996}{}{}.
\newblock
{\BBOQ}\APACrefatitle {The intrinsic {B}ayes factor for model selection and
  prediction} {The intrinsic {B}ayes factor for model selection and
  prediction}.{\BBCQ}
\newblock
\APACjournalVolNumPages{Journal of the American Statistical
  Association}{91}{433}{109--122}.
\newblock
\begin{APACrefDOI} \doi{10.1080/01621459.1996.10476668} \end{APACrefDOI}
\PrintBackRefs{\CurrentBib}

\bibitem [\protect \citeauthoryear {%
Berger%
\ \BBA {} Pericchi%
}{%
Berger%
\ \BBA {} Pericchi%
}{%
{\protect \APACyear {2001}}%
}]{%
bergerpericchi2001}
\APACinsertmetastar {%
bergerpericchi2001}%
\begin{APACrefauthors}%
Berger, J\BPBI O.%
\BCBT {}\ \BBA {} Pericchi, L\BPBI R.%
\end{APACrefauthors}%
\unskip\
\newblock
\APACrefYearMonthDay{2001}{}{}.
\newblock
{\BBOQ}\APACrefatitle {Objective {B}ayesian methods for model selection:
  {I}ntroduction and comparison} {Objective {B}ayesian methods for model
  selection: {I}ntroduction and comparison}.{\BBCQ}
\newblock
\BIn{} P.~Lahiri\ (\BED), \APACrefbtitle {Model selection} {Model selection}\
  (\BVOL~38, \BPGS\ 135--207).
\newblock
\APACaddressPublisher{}{Institute of Mathematical Statistics}.
\newblock
\begin{APACrefDOI} \doi{10.1214/lnms/1215540968} \end{APACrefDOI}
\PrintBackRefs{\CurrentBib}

\bibitem [\protect \citeauthoryear {%
Chipman%
, George%
\BCBL {}\ \BBA {} McCulloch%
}{%
Chipman%
\ \protect \BOthers {.}}{%
{\protect \APACyear {2001}}%
}]{%
chipman2001}
\APACinsertmetastar {%
chipman2001}%
\begin{APACrefauthors}%
Chipman, H.%
, George, E\BPBI I.%
\BCBL {}\ \BBA {} McCulloch, R\BPBI E.%
\end{APACrefauthors}%
\unskip\
\newblock
\APACrefYearMonthDay{2001}{}{}.
\newblock
{\BBOQ}\APACrefatitle {The practical implementation of {B}ayesian model
  selection} {The practical implementation of {B}ayesian model
  selection}.{\BBCQ}
\newblock
\BIn{} \APACrefbtitle {Model selection} {Model selection}\ (\BVOL~38, \BPGS\
  67--116).
\newblock
\APACaddressPublisher{}{Institute of Mathematical Statistics}.
\newblock
\begin{APACrefDOI} \doi{10.1214/lnms/1215540964} \end{APACrefDOI}
\PrintBackRefs{\CurrentBib}

\bibitem [\protect \citeauthoryear {%
Consonni%
, Fouskakis%
, Liseo%
\BCBL {}\ \BBA {} Ntzoufras%
}{%
Consonni%
\ \protect \BOthers {.}}{%
{\protect \APACyear {2018}}%
}]{%
consonni2018}
\APACinsertmetastar {%
consonni2018}%
\begin{APACrefauthors}%
Consonni, G.%
, Fouskakis, D.%
, Liseo, B.%
\BCBL {}\ \BBA {} Ntzoufras, I.%
\end{APACrefauthors}%
\unskip\
\newblock
\APACrefYearMonthDay{2018}{}{}.
\newblock
{\BBOQ}\APACrefatitle {Prior distributions for objective {B}ayesian analysis}
  {Prior distributions for objective {B}ayesian analysis}.{\BBCQ}
\newblock
\APACjournalVolNumPages{Bayesian Analysis}{13}{2}{627--679}.
\newblock
\begin{APACrefDOI} \doi{10.1214/18-BA1103} \end{APACrefDOI}
\PrintBackRefs{\CurrentBib}

\bibitem [\protect \citeauthoryear {%
Crupi%
, Fitelson%
\BCBL {}\ \BBA {} Tentori%
}{%
Crupi%
\ \protect \BOthers {.}}{%
{\protect \APACyear {2008}}%
}]{%
crupi2008}
\APACinsertmetastar {%
crupi2008}%
\begin{APACrefauthors}%
Crupi, V.%
, Fitelson, B.%
\BCBL {}\ \BBA {} Tentori, K.%
\end{APACrefauthors}%
\unskip\
\newblock
\APACrefYearMonthDay{2008}{}{}.
\newblock
{\BBOQ}\APACrefatitle {Probability, confirmation, and the conjunction fallacy}
  {Probability, confirmation, and the conjunction fallacy}.{\BBCQ}
\newblock
\APACjournalVolNumPages{Thinking \& Reasoning}{14}{2}{182--199}.
\newblock
\begin{APACrefDOI} \doi{10.1080/13546780701643406} \end{APACrefDOI}
\PrintBackRefs{\CurrentBib}

\bibitem [\protect \citeauthoryear {%
Dittrich%
, Leenders%
\BCBL {}\ \BBA {} Mulder%
}{%
Dittrich%
\ \protect \BOthers {.}}{%
{\protect \APACyear {2020}}%
}]{%
dittrich2020network}
\APACinsertmetastar {%
dittrich2020network}%
\begin{APACrefauthors}%
Dittrich, D.%
, Leenders, R\BPBI T\BPBI A.%
\BCBL {}\ \BBA {} Mulder, J.%
\end{APACrefauthors}%
\unskip\
\newblock
\APACrefYearMonthDay{2020}{}{}.
\newblock
{\BBOQ}\APACrefatitle {Network autocorrelation modeling: Bayesian techniques
  for estimating and testing multiple network autocorrelations} {Network
  autocorrelation modeling: Bayesian techniques for estimating and testing
  multiple network autocorrelations}.{\BBCQ}
\newblock
\APACjournalVolNumPages{Sociological Methodology}{50}{1}{168--214}.
\PrintBackRefs{\CurrentBib}

\bibitem [\protect \citeauthoryear {%
Fossaluza%
, Izbicki%
, da Silva%
\BCBL {}\ \BBA {} Esteves%
}{%
Fossaluza%
\ \protect \BOthers {.}}{%
{\protect \APACyear {2017}}%
}]{%
fossaluza2017}
\APACinsertmetastar {%
fossaluza2017}%
\begin{APACrefauthors}%
Fossaluza, V.%
, Izbicki, R.%
, da Silva, G\BPBI M.%
\BCBL {}\ \BBA {} Esteves, L\BPBI G.%
\end{APACrefauthors}%
\unskip\
\newblock
\APACrefYearMonthDay{2017}{}{}.
\newblock
{\BBOQ}\APACrefatitle {Coherent hypothesis testing} {Coherent hypothesis
  testing}.{\BBCQ}
\newblock
\APACjournalVolNumPages{The American Statistician}{71}{3}{242--248}.
\newblock
\begin{APACrefDOI} \doi{10.1080/00031305.2016.1237893} \end{APACrefDOI}
\PrintBackRefs{\CurrentBib}

\bibitem [\protect \citeauthoryear {%
Gabriel%
}{%
Gabriel%
}{%
{\protect \APACyear {1969}}%
}]{%
gabriel1969}
\APACinsertmetastar {%
gabriel1969}%
\begin{APACrefauthors}%
Gabriel, K\BPBI R.%
\end{APACrefauthors}%
\unskip\
\newblock
\APACrefYearMonthDay{1969}{}{}.
\newblock
{\BBOQ}\APACrefatitle {Simultaneous test procedures---Some theory of multiple
  comparisons} {Simultaneous test procedures---some theory of multiple
  comparisons}.{\BBCQ}
\newblock
\APACjournalVolNumPages{The Annals of Mathematical
  Statistics}{40}{1}{224--250}.
\newblock
\begin{APACrefDOI} \doi{10.1214/aoms/1177697819} \end{APACrefDOI}
\PrintBackRefs{\CurrentBib}

\bibitem [\protect \citeauthoryear {%
Garcia-Donato%
\ \BBA {} Forte%
}{%
Garcia-Donato%
\ \BBA {} Forte%
}{%
{\protect \APACyear {2018}}%
}]{%
garcia2018bayesian}
\APACinsertmetastar {%
garcia2018bayesian}%
\begin{APACrefauthors}%
Garcia-Donato, G.%
\BCBT {}\ \BBA {} Forte, A.%
\end{APACrefauthors}%
\unskip\
\newblock
\APACrefYearMonthDay{2018}{}{}.
\newblock
{\BBOQ}\APACrefatitle {Bayesian testing, variable selection and model averaging
  in linear models using R with BayesVarSel} {Bayesian testing, variable
  selection and model averaging in linear models using r with
  bayesvarsel}.{\BBCQ}
\newblock
\APACjournalVolNumPages{The {R} Journal,}{}{}{}.
\PrintBackRefs{\CurrentBib}

\bibitem [\protect \citeauthoryear {%
George%
}{%
George%
}{%
{\protect \APACyear {2010}}%
}]{%
george2010}
\APACinsertmetastar {%
george2010}%
\begin{APACrefauthors}%
George, E\BPBI I.%
\end{APACrefauthors}%
\unskip\
\newblock
\APACrefYearMonthDay{2010}{}{}.
\newblock
{\BBOQ}\APACrefatitle {Dilution priors: {C}ompensating for model space
  redundancy} {Dilution priors: {C}ompensating for model space
  redundancy}.{\BBCQ}
\newblock
\BIn{} \APACrefbtitle {Borrowing strength: {T}heory powering applications --
  {A} {F}estschrift for {L}awrence {D}. {B}rown} {Borrowing strength: {T}heory
  powering applications -- {A} {F}estschrift for {L}awrence {D}. {B}rown}\
  (\BVOL~6, \BPGS\ 158--165).
\newblock
\APACaddressPublisher{}{Institute of Mathematical Statistics}.
\PrintBackRefs{\CurrentBib}

\bibitem [\protect \citeauthoryear {%
George%
\ \BBA {} McCulloch%
}{%
George%
\ \BBA {} McCulloch%
}{%
{\protect \APACyear {1993}}%
}]{%
georgemcculloch1993}
\APACinsertmetastar {%
georgemcculloch1993}%
\begin{APACrefauthors}%
George, E\BPBI I.%
\BCBT {}\ \BBA {} McCulloch, R\BPBI E.%
\end{APACrefauthors}%
\unskip\
\newblock
\APACrefYearMonthDay{1993}{}{}.
\newblock
{\BBOQ}\APACrefatitle {Variable selection via {G}ibbs sampling} {Variable
  selection via {G}ibbs sampling}.{\BBCQ}
\newblock
\APACjournalVolNumPages{Journal of the American Statistical
  Association}{88}{423}{881--889}.
\newblock
\begin{APACrefDOI} \doi{10.1080/01621459.1993.10476353} \end{APACrefDOI}
\PrintBackRefs{\CurrentBib}

\bibitem [\protect \citeauthoryear {%
Gr{\"u}nwald%
, de Heide%
\BCBL {}\ \BBA {} Koolen%
}{%
Gr{\"u}nwald%
\ \protect \BOthers {.}}{%
{\protect \APACyear {2024}}%
}]{%
grunwald2024}
\APACinsertmetastar {%
grunwald2024}%
\begin{APACrefauthors}%
Gr{\"u}nwald, P.%
, de Heide, R.%
\BCBL {}\ \BBA {} Koolen, W.%
\end{APACrefauthors}%
\unskip\
\newblock
\APACrefYearMonthDay{2024}{}{}.
\newblock
{\BBOQ}\APACrefatitle {Safe testing} {Safe testing}.{\BBCQ}
\newblock
\APACjournalVolNumPages{Journal of the Royal Statistical Society Series B:
  Statistical Methodology}{86}{5}{1091--1128}.
\newblock
\begin{APACrefDOI} \doi{10.1093/jrsssb/qkae011} \end{APACrefDOI}
\PrintBackRefs{\CurrentBib}

\bibitem [\protect \citeauthoryear {%
Gu%
, Hoijtink%
\BCBL {}\ \BBA {} Mulder%
}{%
Gu%
\ \protect \BOthers {.}}{%
{\protect \APACyear {2022}}%
}]{%
gu2022}
\APACinsertmetastar {%
gu2022}%
\begin{APACrefauthors}%
Gu, X.%
, Hoijtink, H.%
\BCBL {}\ \BBA {} Mulder, J.%
\end{APACrefauthors}%
\unskip\
\newblock
\APACrefYearMonthDay{2022}{}{}.
\newblock
{\BBOQ}\APACrefatitle {Bayesian one-sided variable selection} {Bayesian
  one-sided variable selection}.{\BBCQ}
\newblock
\APACjournalVolNumPages{Multivariate Behavioral Research}{57}{2--3}{264--278}.
\newblock
\begin{APACrefDOI} \doi{10.1080/00273171.2020.1813067} \end{APACrefDOI}
\PrintBackRefs{\CurrentBib}

\bibitem [\protect \citeauthoryear {%
Gu%
, Hoijtink%
, Mulder%
\BCBL {}\ \BBA {} Rosseel%
}{%
Gu%
\ \protect \BOthers {.}}{%
{\protect \APACyear {2019}}%
}]{%
gu2019bain}
\APACinsertmetastar {%
gu2019bain}%
\begin{APACrefauthors}%
Gu, X.%
, Hoijtink, H.%
, Mulder, J.%
\BCBL {}\ \BBA {} Rosseel, Y.%
\end{APACrefauthors}%
\unskip\
\newblock
\APACrefYearMonthDay{2019}{}{}.
\newblock
{\BBOQ}\APACrefatitle {Bain: A program for Bayesian testing of order
  constrained hypotheses in structural equation models} {Bain: A program for
  bayesian testing of order constrained hypotheses in structural equation
  models}.{\BBCQ}
\newblock
\APACjournalVolNumPages{Journal of Statistical Computation and
  Simulation}{89}{8}{1526--1553}.
\PrintBackRefs{\CurrentBib}

\bibitem [\protect \citeauthoryear {%
Gu%
, Mulder%
\BCBL {}\ \BBA {} Hoijtink%
}{%
Gu%
\ \protect \BOthers {.}}{%
{\protect \APACyear {2018}}%
}]{%
gu2018}
\APACinsertmetastar {%
gu2018}%
\begin{APACrefauthors}%
Gu, X.%
, Mulder, J.%
\BCBL {}\ \BBA {} Hoijtink, H.%
\end{APACrefauthors}%
\unskip\
\newblock
\APACrefYearMonthDay{2018}{}{}.
\newblock
{\BBOQ}\APACrefatitle {Approximated adjusted fractional {B}ayes factors: {A}
  general method for testing informative hypotheses} {Approximated adjusted
  fractional {B}ayes factors: {A} general method for testing informative
  hypotheses}.{\BBCQ}
\newblock
\APACjournalVolNumPages{British Journal of Mathematical and Statistical
  Psychology}{71}{2}{229--261}.
\newblock
\begin{APACrefDOI} \doi{10.1111/bmsp.12110} \end{APACrefDOI}
\PrintBackRefs{\CurrentBib}

\bibitem [\protect \citeauthoryear {%
Heck%
\ \BBA {} Davis-Stober%
}{%
Heck%
\ \BBA {} Davis-Stober%
}{%
{\protect \APACyear {2019}}%
}]{%
heck2019}
\APACinsertmetastar {%
heck2019}%
\begin{APACrefauthors}%
Heck, D\BPBI W.%
\BCBT {}\ \BBA {} Davis-Stober, C\BPBI P.%
\end{APACrefauthors}%
\unskip\
\newblock
\APACrefYearMonthDay{2019}{}{}.
\newblock
{\BBOQ}\APACrefatitle {Multinomial models with linear inequality constraints:
  {O}verview and improvements of computational methods for {B}ayesian
  inference} {Multinomial models with linear inequality constraints: {O}verview
  and improvements of computational methods for {B}ayesian inference}.{\BBCQ}
\newblock
\APACjournalVolNumPages{Journal of Mathematical Psychology}{91}{}{70--87}.
\newblock
\begin{APACrefDOI} \doi{10.1016/j.jmp.2019.03.004} \end{APACrefDOI}
\PrintBackRefs{\CurrentBib}

\bibitem [\protect \citeauthoryear {%
Hoijtink%
}{%
Hoijtink%
}{%
{\protect \APACyear {2012}}%
}]{%
hoijtink2012}
\APACinsertmetastar {%
hoijtink2012}%
\begin{APACrefauthors}%
Hoijtink, H.%
\end{APACrefauthors}%
\unskip\
\newblock
\APACrefYear{2012}.
\newblock
\APACrefbtitle {Informative hypotheses: {T}heory and practice for behavioral
  and social scientists} {Informative hypotheses: {T}heory and practice for
  behavioral and social scientists}.
\newblock
\APACaddressPublisher{}{Chapman \& Hall/CRC}.
\PrintBackRefs{\CurrentBib}

\bibitem [\protect \citeauthoryear {%
Hoijtink%
}{%
Hoijtink%
}{%
{\protect \APACyear {2013}}%
}]{%
hoijtink2013}
\APACinsertmetastar {%
hoijtink2013}%
\begin{APACrefauthors}%
Hoijtink, H.%
\end{APACrefauthors}%
\unskip\
\newblock
\APACrefYearMonthDay{2013}{}{}.
\newblock
{\BBOQ}\APACrefatitle {Objective {B}ayes factors for inequality constrained
  hypotheses} {Objective {B}ayes factors for inequality constrained
  hypotheses}.{\BBCQ}
\newblock
\APACjournalVolNumPages{International Statistical Review}{81}{2}{207--229}.
\newblock
\begin{APACrefDOI} \doi{10.1111/insr.12010} \end{APACrefDOI}
\PrintBackRefs{\CurrentBib}

\bibitem [\protect \citeauthoryear {%
Hoijtink%
, Mulder%
, van Lissa%
\BCBL {}\ \BBA {} Gu%
}{%
Hoijtink%
, Mulder%
, van Lissa%
\BCBL {}\ \BBA {} Gu%
}{%
{\protect \APACyear {2019}}%
}]{%
hoijtink2019}
\APACinsertmetastar {%
hoijtink2019}%
\begin{APACrefauthors}%
Hoijtink, H.%
, Mulder, J.%
, van Lissa, C.%
\BCBL {}\ \BBA {} Gu, X.%
\end{APACrefauthors}%
\unskip\
\newblock
\APACrefYearMonthDay{2019}{}{}.
\newblock
{\BBOQ}\APACrefatitle {A tutorial on testing hypotheses using the {B}ayes
  factor} {A tutorial on testing hypotheses using the {B}ayes factor}.{\BBCQ}
\newblock
\APACjournalVolNumPages{Psychological Methods}{24}{5}{539--556}.
\newblock
\begin{APACrefDOI} \doi{10.1037/met0000201} \end{APACrefDOI}
\PrintBackRefs{\CurrentBib}

\bibitem [\protect \citeauthoryear {%
Hoijtink%
, Mulder%
, Van~Lissa%
\BCBL {}\ \BBA {} Gu%
}{%
Hoijtink%
, Mulder%
, Van~Lissa%
\BCBL {}\ \BBA {} Gu%
}{%
{\protect \APACyear {2019}}%
}]{%
hoijtink2019tutorial}
\APACinsertmetastar {%
hoijtink2019tutorial}%
\begin{APACrefauthors}%
Hoijtink, H.%
, Mulder, J.%
, Van~Lissa, C.%
\BCBL {}\ \BBA {} Gu, X.%
\end{APACrefauthors}%
\unskip\
\newblock
\APACrefYearMonthDay{2019}{}{}.
\newblock
{\BBOQ}\APACrefatitle {A tutorial on testing hypotheses using the Bayes
  factor.} {A tutorial on testing hypotheses using the bayes factor.}{\BBCQ}
\newblock
\APACjournalVolNumPages{Psychological methods}{24}{5}{539}.
\PrintBackRefs{\CurrentBib}

\bibitem [\protect \citeauthoryear {%
Ishwaran%
\ \BBA {} Rao%
}{%
Ishwaran%
\ \BBA {} Rao%
}{%
{\protect \APACyear {2005}}%
}]{%
ishwaran2005spike}
\APACinsertmetastar {%
ishwaran2005spike}%
\begin{APACrefauthors}%
Ishwaran, H.%
\BCBT {}\ \BBA {} Rao, J\BPBI S.%
\end{APACrefauthors}%
\unskip\
\newblock
\APACrefYearMonthDay{2005}{}{}.
\newblock
{\BBOQ}\APACrefatitle {Spike and slab variable selection: frequentist and
  Bayesian strategies} {Spike and slab variable selection: frequentist and
  bayesian strategies}.{\BBCQ}
\newblock

\PrintBackRefs{\CurrentBib}

\bibitem [\protect \citeauthoryear {%
Izbicki%
\ \BBA {} Esteves%
}{%
Izbicki%
\ \BBA {} Esteves%
}{%
{\protect \APACyear {2015}}%
}]{%
izbicki2015}
\APACinsertmetastar {%
izbicki2015}%
\begin{APACrefauthors}%
Izbicki, R.%
\BCBT {}\ \BBA {} Esteves, L\BPBI G.%
\end{APACrefauthors}%
\unskip\
\newblock
\APACrefYearMonthDay{2015}{}{}.
\newblock
{\BBOQ}\APACrefatitle {Logical consistency in simultaneous statistical test
  procedures} {Logical consistency in simultaneous statistical test
  procedures}.{\BBCQ}
\newblock
\APACjournalVolNumPages{Logic Journal of the IGPL}{23}{5}{732--758}.
\newblock
\begin{APACrefDOI} \doi{10.1093/jigpal/jzv027} \end{APACrefDOI}
\PrintBackRefs{\CurrentBib}

\bibitem [\protect \citeauthoryear {%
Jeffreys%
}{%
Jeffreys%
}{%
{\protect \APACyear {1961}}%
}]{%
jeffreys1961}
\APACinsertmetastar {%
jeffreys1961}%
\begin{APACrefauthors}%
Jeffreys, H.%
\end{APACrefauthors}%
\unskip\
\newblock
\APACrefYear{1961}.
\newblock
\APACrefbtitle {Theory of probability} {Theory of probability}\
  (\PrintOrdinal{3}\ \BEd).
\newblock
\APACaddressPublisher{}{Oxford University Press}.
\PrintBackRefs{\CurrentBib}

\bibitem [\protect \citeauthoryear {%
Johnson%
, Payne%
, Wang%
, Asher%
\BCBL {}\ \BBA {} Mandal%
}{%
Johnson%
\ \protect \BOthers {.}}{%
{\protect \APACyear {2017}}%
}]{%
Johnson2017}
\APACinsertmetastar {%
Johnson2017}%
\begin{APACrefauthors}%
Johnson, V\BPBI E.%
, Payne, R\BPBI D.%
, Wang, T.%
, Asher, A.%
\BCBL {}\ \BBA {} Mandal, S.%
\end{APACrefauthors}%
\unskip\
\newblock
\APACrefYearMonthDay{2017}{}{}.
\newblock
{\BBOQ}\APACrefatitle {On the reproducibility of psychological science} {On the
  reproducibility of psychological science}.{\BBCQ}
\newblock
\APACjournalVolNumPages{Journal of the American Statistical
  Association}{112}{517}{1--10}.
\newblock
\begin{APACrefDOI} \doi{10.1080/01621459.2016.1240079} \end{APACrefDOI}
\PrintBackRefs{\CurrentBib}

\bibitem [\protect \citeauthoryear {%
Kass%
\ \BBA {} Raftery%
}{%
Kass%
\ \BBA {} Raftery%
}{%
{\protect \APACyear {1995}}%
}]{%
kassraftery1995}
\APACinsertmetastar {%
kassraftery1995}%
\begin{APACrefauthors}%
Kass, R\BPBI E.%
\BCBT {}\ \BBA {} Raftery, A\BPBI E.%
\end{APACrefauthors}%
\unskip\
\newblock
\APACrefYearMonthDay{1995}{}{}.
\newblock
{\BBOQ}\APACrefatitle {Bayes factors} {Bayes factors}.{\BBCQ}
\newblock
\APACjournalVolNumPages{Journal of the American Statistical
  Association}{90}{430}{773--795}.
\newblock
\begin{APACrefDOI} \doi{10.1080/01621459.1995.10476572} \end{APACrefDOI}
\PrintBackRefs{\CurrentBib}

\bibitem [\protect \citeauthoryear {%
Klaassen%
}{%
Klaassen%
}{%
{\protect \APACyear {2020}}%
}]{%
klaassen2020}
\APACinsertmetastar {%
klaassen2020}%
\begin{APACrefauthors}%
Klaassen, F.%
\end{APACrefauthors}%
\unskip\
\newblock
\APACrefYear{2020}.
\unskip\
\newblock
\APACrefbtitle {Staying in the loop: {P}rior odds, {B}ayes factor, posterior
  odds} {Staying in the loop: {P}rior odds, {B}ayes factor, posterior odds}\
  \APACtypeAddressSchool {\BUPhD}{}{}.
\unskip\
\newblock
\APACaddressSchool {}{Utrecht University}.
\PrintBackRefs{\CurrentBib}

\bibitem [\protect \citeauthoryear {%
Klugkist%
, Laudy%
\BCBL {}\ \BBA {} Hoijtink%
}{%
Klugkist%
\ \protect \BOthers {.}}{%
{\protect \APACyear {2005}}%
{\protect \APACexlab {{\protect \BCnt {1}}}}}]{%
klugkist2005reply}
\APACinsertmetastar {%
klugkist2005reply}%
\begin{APACrefauthors}%
Klugkist, I.%
, Laudy, O.%
\BCBL {}\ \BBA {} Hoijtink, H.%
\end{APACrefauthors}%
\unskip\
\newblock
\APACrefYearMonthDay{2005{\protect \BCnt {1}}}{}{}.
\newblock
{\BBOQ}\APACrefatitle {Bayesian eggs and {B}ayesian omelettes: {R}eply to
  {S}tern (2005)} {Bayesian eggs and {B}ayesian omelettes: {R}eply to {S}tern
  (2005)}.{\BBCQ}
\newblock
\APACjournalVolNumPages{Psychological Methods}{10}{4}{500--503}.
\newblock
\begin{APACrefDOI} \doi{10.1037/1082-989X.10.4.500} \end{APACrefDOI}
\PrintBackRefs{\CurrentBib}

\bibitem [\protect \citeauthoryear {%
Klugkist%
, Laudy%
\BCBL {}\ \BBA {} Hoijtink%
}{%
Klugkist%
\ \protect \BOthers {.}}{%
{\protect \APACyear {2005}}%
{\protect \APACexlab {{\protect \BCnt {2}}}}}]{%
klugkist2005}
\APACinsertmetastar {%
klugkist2005}%
\begin{APACrefauthors}%
Klugkist, I.%
, Laudy, O.%
\BCBL {}\ \BBA {} Hoijtink, H.%
\end{APACrefauthors}%
\unskip\
\newblock
\APACrefYearMonthDay{2005{\protect \BCnt {2}}}{}{}.
\newblock
{\BBOQ}\APACrefatitle {Inequality constrained analysis of variance: {A}
  {B}ayesian approach} {Inequality constrained analysis of variance: {A}
  {B}ayesian approach}.{\BBCQ}
\newblock
\APACjournalVolNumPages{Psychological Methods}{10}{4}{477--493}.
\newblock
\begin{APACrefDOI} \doi{10.1037/1082-989X.10.4.477} \end{APACrefDOI}
\PrintBackRefs{\CurrentBib}

\bibitem [\protect \citeauthoryear {%
Klugkist%
, Laudy%
\BCBL {}\ \BBA {} Hoijtink%
}{%
Klugkist%
\ \protect \BOthers {.}}{%
{\protect \APACyear {2010}}%
}]{%
Klugkist:2010}
\APACinsertmetastar {%
Klugkist:2010}%
\begin{APACrefauthors}%
Klugkist, I.%
, Laudy, O.%
\BCBL {}\ \BBA {} Hoijtink, H.%
\end{APACrefauthors}%
\unskip\
\newblock
\APACrefYearMonthDay{2010}{}{}.
\newblock
{\BBOQ}\APACrefatitle {{B}ayesian evaluation of inequality and equality
  constrained hypotheses for contingency tables} {{B}ayesian evaluation of
  inequality and equality constrained hypotheses for contingency
  tables}.{\BBCQ}
\newblock
\APACjournalVolNumPages{Psychological Methods}{15}{}{281--299}.
\PrintBackRefs{\CurrentBib}

\bibitem [\protect \citeauthoryear {%
Lavine%
\ \BBA {} Schervish%
}{%
Lavine%
\ \BBA {} Schervish%
}{%
{\protect \APACyear {1999}}%
}]{%
lavineschervish1999}
\APACinsertmetastar {%
lavineschervish1999}%
\begin{APACrefauthors}%
Lavine, M.%
\BCBT {}\ \BBA {} Schervish, M\BPBI J.%
\end{APACrefauthors}%
\unskip\
\newblock
\APACrefYearMonthDay{1999}{}{}.
\newblock
{\BBOQ}\APACrefatitle {Bayes factors: {W}hat they are and what they are not}
  {Bayes factors: {W}hat they are and what they are not}.{\BBCQ}
\newblock
\APACjournalVolNumPages{The American Statistician}{53}{2}{119--122}.
\newblock
\begin{APACrefDOI} \doi{10.1080/00031305.1999.10474443} \end{APACrefDOI}
\PrintBackRefs{\CurrentBib}

\bibitem [\protect \citeauthoryear {%
Liang%
, Paulo%
, Molina%
, Clyde%
\BCBL {}\ \BBA {} Berger%
}{%
Liang%
\ \protect \BOthers {.}}{%
{\protect \APACyear {2008}}%
}]{%
liang2008}
\APACinsertmetastar {%
liang2008}%
\begin{APACrefauthors}%
Liang, F.%
, Paulo, R.%
, Molina, G.%
, Clyde, M\BPBI A.%
\BCBL {}\ \BBA {} Berger, J\BPBI O.%
\end{APACrefauthors}%
\unskip\
\newblock
\APACrefYearMonthDay{2008}{}{}.
\newblock
{\BBOQ}\APACrefatitle {Mixtures of $g$ priors for {B}ayesian variable
  selection} {Mixtures of $g$ priors for {B}ayesian variable selection}.{\BBCQ}
\newblock
\APACjournalVolNumPages{Journal of the American Statistical
  Association}{103}{481}{410--423}.
\newblock
\begin{APACrefDOI} \doi{10.1198/016214507000001337} \end{APACrefDOI}
\PrintBackRefs{\CurrentBib}

\bibitem [\protect \citeauthoryear {%
Marsman%
, Huth%
, Waldorp%
\BCBL {}\ \BBA {} Ntzoufras%
}{%
Marsman%
\ \protect \BOthers {.}}{%
{\protect \APACyear {2022}}%
}]{%
marsman2022}
\APACinsertmetastar {%
marsman2022}%
\begin{APACrefauthors}%
Marsman, M.%
, Huth, K.%
, Waldorp, L\BPBI J.%
\BCBL {}\ \BBA {} Ntzoufras, I.%
\end{APACrefauthors}%
\unskip\
\newblock
\APACrefYearMonthDay{2022}{}{}.
\newblock
{\BBOQ}\APACrefatitle {Objective {B}ayesian edge screening and structure
  selection for {I}sing networks} {Objective {B}ayesian edge screening and
  structure selection for {I}sing networks}.{\BBCQ}
\newblock
\APACjournalVolNumPages{Psychometrika}{87}{1}{47--82}.
\newblock
\begin{APACrefDOI} \doi{10.1007/s11336-022-09848-8} \end{APACrefDOI}
\PrintBackRefs{\CurrentBib}

\bibitem [\protect \citeauthoryear {%
Meehl%
}{%
Meehl%
}{%
{\protect \APACyear {1967}}%
}]{%
Meehl1967}
\APACinsertmetastar {%
Meehl1967}%
\begin{APACrefauthors}%
Meehl, P\BPBI E.%
\end{APACrefauthors}%
\unskip\
\newblock
\APACrefYearMonthDay{1967}{}{}.
\newblock
{\BBOQ}\APACrefatitle {Theory-testing in psychology and physics: {A}
  methodological paradox} {Theory-testing in psychology and physics: {A}
  methodological paradox}.{\BBCQ}
\newblock
\APACjournalVolNumPages{Philosophy of Science}{34}{2}{103--115}.
\newblock
\begin{APACrefDOI} \doi{10.1086/288135} \end{APACrefDOI}
\PrintBackRefs{\CurrentBib}

\bibitem [\protect \citeauthoryear {%
Mohammadi%
\ \BBA {} Wit%
}{%
Mohammadi%
\ \BBA {} Wit%
}{%
{\protect \APACyear {2015}}%
}]{%
mohammadi2015}
\APACinsertmetastar {%
mohammadi2015}%
\begin{APACrefauthors}%
Mohammadi, A.%
\BCBT {}\ \BBA {} Wit, E\BPBI C.%
\end{APACrefauthors}%
\unskip\
\newblock
\APACrefYearMonthDay{2015}{}{}.
\newblock
{\BBOQ}\APACrefatitle {Bayesian structure learning in sparse {G}aussian
  graphical models} {Bayesian structure learning in sparse {G}aussian graphical
  models}.{\BBCQ}
\newblock
\APACjournalVolNumPages{Bayesian Analysis}{10}{1}{109--138}.
\newblock
\begin{APACrefDOI} \doi{10.1214/14-BA889} \end{APACrefDOI}
\PrintBackRefs{\CurrentBib}

\bibitem [\protect \citeauthoryear {%
Mulder%
}{%
Mulder%
}{%
{\protect \APACyear {2014}}%
{\protect \APACexlab {{\protect \BCnt {1}}}}}]{%
mulder2014}
\APACinsertmetastar {%
mulder2014}%
\begin{APACrefauthors}%
Mulder, J.%
\end{APACrefauthors}%
\unskip\
\newblock
\APACrefYearMonthDay{2014{\protect \BCnt {1}}}{}{}.
\newblock
{\BBOQ}\APACrefatitle {Bayes factors for testing inequality constrained
  hypotheses: {I}ssues with prior specification} {Bayes factors for testing
  inequality constrained hypotheses: {I}ssues with prior specification}.{\BBCQ}
\newblock
\APACjournalVolNumPages{British Journal of Mathematical and Statistical
  Psychology}{67}{1}{153--171}.
\newblock
\begin{APACrefDOI} \doi{10.1111/bmsp.12013} \end{APACrefDOI}
\PrintBackRefs{\CurrentBib}

\bibitem [\protect \citeauthoryear {%
Mulder%
}{%
Mulder%
}{%
{\protect \APACyear {2014}}%
{\protect \APACexlab {{\protect \BCnt {2}}}}}]{%
mulder:2014b}
\APACinsertmetastar {%
mulder:2014b}%
\begin{APACrefauthors}%
Mulder, J.%
\end{APACrefauthors}%
\unskip\
\newblock
\APACrefYearMonthDay{2014{\protect \BCnt {2}}}{}{}.
\newblock
{\BBOQ}\APACrefatitle {Prior adjusted default {B}ayes factors for testing
  (in)equality constrained hypotheses} {Prior adjusted default {B}ayes factors
  for testing (in)equality constrained hypotheses}.{\BBCQ}
\newblock
\APACjournalVolNumPages{Computational Statistics \& Data
  Analysis}{71}{}{448--463}.
\newblock
\begin{APACrefDOI} \doi{10.1016/j.csda.2013.07.017} \end{APACrefDOI}
\PrintBackRefs{\CurrentBib}

\bibitem [\protect \citeauthoryear {%
Mulder%
\ \BBA {} Gelissen%
}{%
Mulder%
\ \BBA {} Gelissen%
}{%
{\protect \APACyear {2023}}%
}]{%
mulder2023bayes}
\APACinsertmetastar {%
mulder2023bayes}%
\begin{APACrefauthors}%
Mulder, J.%
\BCBT {}\ \BBA {} Gelissen, J\BPBI P.%
\end{APACrefauthors}%
\unskip\
\newblock
\APACrefYearMonthDay{2023}{}{}.
\newblock
{\BBOQ}\APACrefatitle {Bayes factor testing of equality and order constraints
  on measures of association in social research} {Bayes factor testing of
  equality and order constraints on measures of association in social
  research}.{\BBCQ}
\newblock
\APACjournalVolNumPages{Journal of Applied Statistics}{50}{2}{315--351}.
\PrintBackRefs{\CurrentBib}

\bibitem [\protect \citeauthoryear {%
Mulder%
\ \BBA {} Gu%
}{%
Mulder%
\ \BBA {} Gu%
}{%
{\protect \APACyear {2022}}%
}]{%
mulder2022bayesian}
\APACinsertmetastar {%
mulder2022bayesian}%
\begin{APACrefauthors}%
Mulder, J.%
\BCBT {}\ \BBA {} Gu, X.%
\end{APACrefauthors}%
\unskip\
\newblock
\APACrefYearMonthDay{2022}{}{}.
\newblock
{\BBOQ}\APACrefatitle {Bayesian testing of scientific expectations under
  multivariate normal linear models} {Bayesian testing of scientific
  expectations under multivariate normal linear models}.{\BBCQ}
\newblock
\APACjournalVolNumPages{Multivariate Behavioral Research}{57}{5}{767--783}.
\PrintBackRefs{\CurrentBib}

\bibitem [\protect \citeauthoryear {%
Mulder%
, Hoijtink%
\BCBL {}\ \BBA {} Klugkist%
}{%
Mulder%
\ \protect \BOthers {.}}{%
{\protect \APACyear {2010}}%
}]{%
mulder2010}
\APACinsertmetastar {%
mulder2010}%
\begin{APACrefauthors}%
Mulder, J.%
, Hoijtink, H.%
\BCBL {}\ \BBA {} Klugkist, I.%
\end{APACrefauthors}%
\unskip\
\newblock
\APACrefYearMonthDay{2010}{}{}.
\newblock
{\BBOQ}\APACrefatitle {Equality and Inequality Constrained Multivariate Linear
  Models: Objective Model Selection Using Constrained Posterior Priors}
  {Equality and inequality constrained multivariate linear models: Objective
  model selection using constrained posterior priors}.{\BBCQ}
\newblock
\APACjournalVolNumPages{Journal of Statistical Planning and
  Inference}{140}{}{887--906}.
\PrintBackRefs{\CurrentBib}

\bibitem [\protect \citeauthoryear {%
Mulder%
\ \protect \BOthers {.}}{%
Mulder%
\ \protect \BOthers {.}}{%
{\protect \APACyear {2009}}%
}]{%
mulder2009bayesian}
\APACinsertmetastar {%
mulder2009bayesian}%
\begin{APACrefauthors}%
Mulder, J.%
, Klugkist, I.%
, Van De~Schoot, R.%
, Meeus, W\BPBI H.%
, Selfhout, M.%
\BCBL {}\ \BBA {} Hoijtink, H.%
\end{APACrefauthors}%
\unskip\
\newblock
\APACrefYearMonthDay{2009}{}{}.
\newblock
{\BBOQ}\APACrefatitle {Bayesian model selection of informative hypotheses for
  repeated measurements} {Bayesian model selection of informative hypotheses
  for repeated measurements}.{\BBCQ}
\newblock
\APACjournalVolNumPages{Journal of Mathematical Psychology}{53}{6}{530--546}.
\PrintBackRefs{\CurrentBib}

\bibitem [\protect \citeauthoryear {%
Mulder%
\ \BBA {} Olsson-Collentine%
}{%
Mulder%
\ \BBA {} Olsson-Collentine%
}{%
{\protect \APACyear {2019}}%
}]{%
mulderolsson2019}
\APACinsertmetastar {%
mulderolsson2019}%
\begin{APACrefauthors}%
Mulder, J.%
\BCBT {}\ \BBA {} Olsson-Collentine, A.%
\end{APACrefauthors}%
\unskip\
\newblock
\APACrefYearMonthDay{2019}{}{}.
\newblock
{\BBOQ}\APACrefatitle {Simple {B}ayesian testing of scientific expectations in
  linear regression models} {Simple {B}ayesian testing of scientific
  expectations in linear regression models}.{\BBCQ}
\newblock
\APACjournalVolNumPages{Behavior Research Methods}{51}{3}{1117--1130}.
\newblock
\begin{APACrefDOI} \doi{10.3758/s13428-018-01196-9} \end{APACrefDOI}
\PrintBackRefs{\CurrentBib}

\bibitem [\protect \citeauthoryear {%
Mulder%
\ \BBA {} Raftery%
}{%
Mulder%
\ \BBA {} Raftery%
}{%
{\protect \APACyear {2022}}%
}]{%
mulderraftery2022}
\APACinsertmetastar {%
mulderraftery2022}%
\begin{APACrefauthors}%
Mulder, J.%
\BCBT {}\ \BBA {} Raftery, A\BPBI E.%
\end{APACrefauthors}%
\unskip\
\newblock
\APACrefYearMonthDay{2022}{}{}.
\newblock
{\BBOQ}\APACrefatitle {{BIC} extensions for order-constrained model selection}
  {{BIC} extensions for order-constrained model selection}.{\BBCQ}
\newblock
\APACjournalVolNumPages{Sociological Methods \& Research}{51}{2}{471--498}.
\newblock
\begin{APACrefDOI} \doi{10.1177/0049124119882459} \end{APACrefDOI}
\PrintBackRefs{\CurrentBib}

\bibitem [\protect \citeauthoryear {%
Mulder%
\ \protect \BOthers {.}}{%
Mulder%
\ \protect \BOthers {.}}{%
{\protect \APACyear {2021}}%
}]{%
mulder2021bfpack}
\APACinsertmetastar {%
mulder2021bfpack}%
\begin{APACrefauthors}%
Mulder, J.%
, Williams, D\BPBI R.%
, Gu, X.%
, Tomarken, A.%
, B{\"o}ing-Messing, F.%
, Olsson-Collentine, A.%
\BDBL {}van Lissa, C.%
\end{APACrefauthors}%
\unskip\
\newblock
\APACrefYearMonthDay{2021}{}{}.
\newblock
{\BBOQ}\APACrefatitle {{BFpack}: {F}lexible {B}ayes factor testing of
  scientific theories in {R}} {{BFpack}: {F}lexible {B}ayes factor testing of
  scientific theories in {R}}.{\BBCQ}
\newblock
\APACjournalVolNumPages{Journal of Statistical Software}{100}{18}{1--63}.
\newblock
\begin{APACrefDOI} \doi{10.18637/jss.v100.i18} \end{APACrefDOI}
\PrintBackRefs{\CurrentBib}

\bibitem [\protect \citeauthoryear {%
O'Hagan%
}{%
O'Hagan%
}{%
{\protect \APACyear {1995}}%
}]{%
ohagan1995}
\APACinsertmetastar {%
ohagan1995}%
\begin{APACrefauthors}%
O'Hagan, A.%
\end{APACrefauthors}%
\unskip\
\newblock
\APACrefYearMonthDay{1995}{}{}.
\newblock
{\BBOQ}\APACrefatitle {Fractional {B}ayes factors for model comparison}
  {Fractional {B}ayes factors for model comparison}.{\BBCQ}
\newblock
\APACjournalVolNumPages{Journal of the Royal Statistical Society: Series B
  (Methodological)}{57}{1}{99--138}.
\newblock
\begin{APACrefDOI} \doi{10.1111/j.2517-6161.1995.tb02017.x} \end{APACrefDOI}
\PrintBackRefs{\CurrentBib}

\bibitem [\protect \citeauthoryear {%
Rouder%
\ \BBA {} Morey%
}{%
Rouder%
\ \BBA {} Morey%
}{%
{\protect \APACyear {2012}}%
}]{%
roudermorey2012}
\APACinsertmetastar {%
roudermorey2012}%
\begin{APACrefauthors}%
Rouder, J\BPBI N.%
\BCBT {}\ \BBA {} Morey, R\BPBI D.%
\end{APACrefauthors}%
\unskip\
\newblock
\APACrefYearMonthDay{2012}{}{}.
\newblock
{\BBOQ}\APACrefatitle {Default {B}ayes factors for model selection in
  regression} {Default {B}ayes factors for model selection in
  regression}.{\BBCQ}
\newblock
\APACjournalVolNumPages{Multivariate Behavioral Research}{47}{6}{877--903}.
\newblock
\begin{APACrefDOI} \doi{10.1080/00273171.2012.734737} \end{APACrefDOI}
\PrintBackRefs{\CurrentBib}

\bibitem [\protect \citeauthoryear {%
Sarafoglou%
\ \protect \BOthers {.}}{%
Sarafoglou%
\ \protect \BOthers {.}}{%
{\protect \APACyear {2023}}%
}]{%
sarafoglou2023}
\APACinsertmetastar {%
sarafoglou2023}%
\begin{APACrefauthors}%
Sarafoglou, A.%
, Haaf, J\BPBI M.%
, Ly, A.%
, Gronau, Q\BPBI F.%
, Wagenmakers, E\BHBI J.%
\BCBL {}\ \BBA {} Marsman, M.%
\end{APACrefauthors}%
\unskip\
\newblock
\APACrefYearMonthDay{2023}{}{}.
\newblock
{\BBOQ}\APACrefatitle {Evaluating multinomial order restrictions with bridge
  sampling} {Evaluating multinomial order restrictions with bridge
  sampling}.{\BBCQ}
\newblock
\APACjournalVolNumPages{Psychological Methods}{28}{2}{322--338}.
\newblock
\begin{APACrefDOI} \doi{10.1037/met0000411} \end{APACrefDOI}
\PrintBackRefs{\CurrentBib}

\bibitem [\protect \citeauthoryear {%
Scott%
\ \BBA {} Berger%
}{%
Scott%
\ \BBA {} Berger%
}{%
{\protect \APACyear {2010}}%
}]{%
scottberger2010}
\APACinsertmetastar {%
scottberger2010}%
\begin{APACrefauthors}%
Scott, J\BPBI G.%
\BCBT {}\ \BBA {} Berger, J\BPBI O.%
\end{APACrefauthors}%
\unskip\
\newblock
\APACrefYearMonthDay{2010}{}{}.
\newblock
{\BBOQ}\APACrefatitle {Bayes and empirical-{B}ayes multiplicity adjustment in
  the variable-selection problem} {Bayes and empirical-{B}ayes multiplicity
  adjustment in the variable-selection problem}.{\BBCQ}
\newblock
\APACjournalVolNumPages{The Annals of Statistics}{38}{5}{2587--2619}.
\newblock
\begin{APACrefDOI} \doi{10.1214/10-AOS792} \end{APACrefDOI}
\PrintBackRefs{\CurrentBib}

\bibitem [\protect \citeauthoryear {%
Stern%
}{%
Stern%
}{%
{\protect \APACyear {2005}}%
}]{%
stern2005}
\APACinsertmetastar {%
stern2005}%
\begin{APACrefauthors}%
Stern, H\BPBI S.%
\end{APACrefauthors}%
\unskip\
\newblock
\APACrefYearMonthDay{2005}{}{}.
\newblock
{\BBOQ}\APACrefatitle {Model inference or model selection: {D}iscussion of
  {K}lugkist, {L}audy, and {H}oijtink (2005)} {Model inference or model
  selection: {D}iscussion of {K}lugkist, {L}audy, and {H}oijtink
  (2005)}.{\BBCQ}
\newblock
\APACjournalVolNumPages{Psychological Methods}{10}{4}{494--499}.
\newblock
\begin{APACrefDOI} \doi{10.1037/1082-989X.10.4.494} \end{APACrefDOI}
\PrintBackRefs{\CurrentBib}

\bibitem [\protect \citeauthoryear {%
Tentori%
, Crupi%
\BCBL {}\ \BBA {} Russo%
}{%
Tentori%
\ \protect \BOthers {.}}{%
{\protect \APACyear {2013}}%
}]{%
tentori2013}
\APACinsertmetastar {%
tentori2013}%
\begin{APACrefauthors}%
Tentori, K.%
, Crupi, V.%
\BCBL {}\ \BBA {} Russo, S.%
\end{APACrefauthors}%
\unskip\
\newblock
\APACrefYearMonthDay{2013}{}{}.
\newblock
{\BBOQ}\APACrefatitle {On the determinants of the conjunction fallacy:
  {P}robability versus inductive confirmation} {On the determinants of the
  conjunction fallacy: {P}robability versus inductive confirmation}.{\BBCQ}
\newblock
\APACjournalVolNumPages{Journal of Experimental Psychology:
  General}{142}{1}{235--255}.
\newblock
\begin{APACrefDOI} \doi{10.1037/a0028770} \end{APACrefDOI}
\PrintBackRefs{\CurrentBib}

\bibitem [\protect \citeauthoryear {%
Tversky%
\ \BBA {} Kahneman%
}{%
Tversky%
\ \BBA {} Kahneman%
}{%
{\protect \APACyear {1983}}%
}]{%
tverskykahneman1983}
\APACinsertmetastar {%
tverskykahneman1983}%
\begin{APACrefauthors}%
Tversky, A.%
\BCBT {}\ \BBA {} Kahneman, D.%
\end{APACrefauthors}%
\unskip\
\newblock
\APACrefYearMonthDay{1983}{}{}.
\newblock
{\BBOQ}\APACrefatitle {Extensional versus intuitive reasoning: {T}he
  conjunction fallacy in probability judgment} {Extensional versus intuitive
  reasoning: {T}he conjunction fallacy in probability judgment}.{\BBCQ}
\newblock
\APACjournalVolNumPages{Psychological Review}{90}{4}{293--315}.
\newblock
\begin{APACrefDOI} \doi{10.1037/0033-295X.90.4.293} \end{APACrefDOI}
\PrintBackRefs{\CurrentBib}

\bibitem [\protect \citeauthoryear {%
van~den Bergh%
\ \BBA {} Dablander%
}{%
van~den Bergh%
\ \BBA {} Dablander%
}{%
{\protect \APACyear {2026}}%
}]{%
vandenbergh2026}
\APACinsertmetastar {%
vandenbergh2026}%
\begin{APACrefauthors}%
van~den Bergh, D.%
\BCBT {}\ \BBA {} Dablander, F.%
\end{APACrefauthors}%
\unskip\
\newblock
\APACrefYearMonthDay{2026}{}{}.
\newblock
{\BBOQ}\APACrefatitle {Flexible {B}ayesian multiple comparison adjustment using
  {D}irichlet process and beta-binomial model priors} {Flexible {B}ayesian
  multiple comparison adjustment using {D}irichlet process and beta-binomial
  model priors}.{\BBCQ}
\newblock
\APACjournalVolNumPages{The American Statistician}{80}{}{217--231}.
\newblock
\begin{APACrefDOI} \doi{10.1080/00031305.2025.2561146} \end{APACrefDOI}
\PrintBackRefs{\CurrentBib}

\bibitem [\protect \citeauthoryear {%
van~den Bergh%
\ \protect \BOthers {.}}{%
van~den Bergh%
\ \protect \BOthers {.}}{%
{\protect \APACyear {2020}}%
}]{%
van2020tutorial}
\APACinsertmetastar {%
van2020tutorial}%
\begin{APACrefauthors}%
van~den Bergh, D.%
, Van~Doorn, J.%
, Marsman, M.%
, Draws, T.%
, Van~Kesteren, E\BHBI J.%
, Derks, K.%
\BDBL {}others%
\end{APACrefauthors}%
\unskip\
\newblock
\APACrefYearMonthDay{2020}{}{}.
\newblock
{\BBOQ}\APACrefatitle {A tutorial on conducting and interpreting a Bayesian
  ANOVA in JASP} {A tutorial on conducting and interpreting a bayesian anova in
  jasp}.{\BBCQ}
\newblock
\APACjournalVolNumPages{L’Ann{\'e}e psychologique}{120}{1}{73--96}.
\PrintBackRefs{\CurrentBib}

\bibitem [\protect \citeauthoryear {%
Van~Lissa%
\ \protect \BOthers {.}}{%
Van~Lissa%
\ \protect \BOthers {.}}{%
{\protect \APACyear {2021}}%
}]{%
vanlissa2021}
\APACinsertmetastar {%
vanlissa2021}%
\begin{APACrefauthors}%
Van~Lissa, C\BPBI J.%
, Gu, X.%
, Mulder, J.%
, Rosseel, Y.%
, Van~Zundert, C.%
\BCBL {}\ \BBA {} Hoijtink, H.%
\end{APACrefauthors}%
\unskip\
\newblock
\APACrefYearMonthDay{2021}{}{}.
\newblock
{\BBOQ}\APACrefatitle {Teacher's Corner: Evaluating Informative Hypotheses
  Using the Bayes Factor in Structural Equation Models} {Teacher's corner:
  Evaluating informative hypotheses using the bayes factor in structural
  equation models}.{\BBCQ}
\newblock
\APACjournalVolNumPages{Structural Equation Modeling: A Multidisciplinary
  Journal}{28}{2}{292--301}.
\PrintBackRefs{\CurrentBib}

\bibitem [\protect \citeauthoryear {%
Williams%
\ \BBA {} Mulder%
}{%
Williams%
\ \BBA {} Mulder%
}{%
{\protect \APACyear {2020}}%
}]{%
williamsmulder2020}
\APACinsertmetastar {%
williamsmulder2020}%
\begin{APACrefauthors}%
Williams, D\BPBI R.%
\BCBT {}\ \BBA {} Mulder, J.%
\end{APACrefauthors}%
\unskip\
\newblock
\APACrefYearMonthDay{2020}{}{}.
\newblock
{\BBOQ}\APACrefatitle {Bayesian hypothesis testing for {G}aussian graphical
  models: {C}onditional independence and order constraints} {Bayesian
  hypothesis testing for {G}aussian graphical models: {C}onditional
  independence and order constraints}.{\BBCQ}
\newblock
\APACjournalVolNumPages{Journal of Mathematical Psychology}{99}{}{Article
  102441}.
\newblock
\begin{APACrefDOI} \doi{10.1016/j.jmp.2020.102441} \end{APACrefDOI}
\PrintBackRefs{\CurrentBib}

\bibitem [\protect \citeauthoryear {%
Zellner%
}{%
Zellner%
}{%
{\protect \APACyear {1986}}%
}]{%
zellner1986}
\APACinsertmetastar {%
zellner1986}%
\begin{APACrefauthors}%
Zellner, A.%
\end{APACrefauthors}%
\unskip\
\newblock
\APACrefYearMonthDay{1986}{}{}.
\newblock
{\BBOQ}\APACrefatitle {On assessing prior distributions and {B}ayesian
  regression analysis with $g$-prior distributions} {On assessing prior
  distributions and {B}ayesian regression analysis with $g$-prior
  distributions}.{\BBCQ}
\newblock
\BIn{} P\BPBI K.~Goel\ \BBA {} A.~Zellner\ (\BEDS), \APACrefbtitle {Bayesian
  inference and decision techniques: {E}ssays in honor of {B}runo de {F}inetti}
  {Bayesian inference and decision techniques: {E}ssays in honor of {B}runo de
  {F}inetti}\ (\BPGS\ 233--243).
\newblock
\APACaddressPublisher{}{North-Holland}.
\PrintBackRefs{\CurrentBib}

\bibitem [\protect \citeauthoryear {%
Zellner%
\ \BBA {} Siow%
}{%
Zellner%
\ \BBA {} Siow%
}{%
{\protect \APACyear {1980}}%
}]{%
zellnersiow1980}
\APACinsertmetastar {%
zellnersiow1980}%
\begin{APACrefauthors}%
Zellner, A.%
\BCBT {}\ \BBA {} Siow, A.%
\end{APACrefauthors}%
\unskip\
\newblock
\APACrefYearMonthDay{1980}{}{}.
\newblock
{\BBOQ}\APACrefatitle {Posterior odds ratios for selected regression
  hypotheses} {Posterior odds ratios for selected regression
  hypotheses}.{\BBCQ}
\newblock
\BIn{} J\BPBI M.~Bernardo, M\BPBI H.~DeGroot, D\BPBI V.~Lindley\BCBL {}\ \BBA
  {} A\BPBI F\BPBI M.~Smith\ (\BEDS), \APACrefbtitle {Bayesian statistics}
  {Bayesian statistics}\ (\BPGS\ 585--603).
\newblock
\APACaddressPublisher{}{Valencia University Press}.
\PrintBackRefs{\CurrentBib}

\end{thebibliography}

\appendix
\section{Proofs}\label{app:proofs}

\begin{proof}[Proof of Lemma~\ref{lem:post}]
Using \eqref{eq:BF}, $m_{s,j}(\Y)=m(\Y\,|\,\mathcal{M}_s)f_{s,j}(\Y)/c_{s,j}$, so that
\[
m_t(\Y)=\sum_{(s,j)\in\mathcal{J}_t}\frac{P(\mathcal{M}_s)c_{s,j}}{\Pi_t}\,m_{s,j}(\Y)
=\frac{m(\Y)\,\Pi_t(\Y)}{\Pi_t},
\]
with $m(\Y)=\sum_{s\in\mathcal{S}}P(\mathcal{M}_s)m(\Y\,|\,\mathcal{M}_s)$. Bayes theorem with prior
probabilities proportional to $\Pi_t$ gives \eqref{eq:postcomp}. When the hypotheses are exhaustive,
the relative fits of the included regions sum to one within each equality structure, so that
$\sum_t\Pi_t(\Y)=\sum_{s}P(\mathcal{M}_s|\Y)=1$. Under the uniform prior, the posterior probabilities
are proportional to $m_{s,j}(\Y)$, which gives \eqref{eq:postunif}.
\end{proof}

\begin{proof}[Proof of Proposition~\ref{prop:inv}]
In \eqref{eq:postcomp}, the numerator $\Pi_t(\Y)$ only depends on the pairs in $\mathcal{J}_t$, and
$P(\mathcal{M}_s|\Y)$ does not depend on the hypothesis space. Because relative fits are additive over
disjoint regions, the denominator equals
$\sum_{s\in\mathcal{S}}P(\mathcal{M}_s|\Y)\Pr(\bth\in U_s\,|\,\mathcal{M}_s,\Y)$, with $U_s$ the union
of the regions under $\mathcal{M}_s$ that are included in the hypotheses, which is the same for both
hypothesis spaces. This gives (i), (ii) follows from the additivity of $\Pi_t(\Y)$, and (iii) from
$\Pi_t(\Y)\leq\Pi_{t'}(\Y)$ as all terms are nonnegative.
\end{proof}

\section{R Code for the Empirical Illustration}\label{app:code}

{\scriptsize
\begin{verbatim}
# ---------------------------------------------------------------------------
# Empirical illustration
#
# For a pair of effects (b1, b2):
#   H1: b1 = b2 = 0 ,  H2: b1 > b2 > 0 ,  H3: complement
#   H1*: b1 = b2 = 0 , H2*: not H1*
#
# Carlsson and Sinclair:  feminist > stigma > 0
# EVS Germany:            education > class > 0
#
# Posterior probabilities under the uniform and the complexity-induced prior,
# using the AFBF in BFpack and the JZS Bayes factor in BayesFactor.
# ---------------------------------------------------------------------------

library(BFpack)
library(BayesFactor)
library(mvtnorm)
library(haven)
library(dplyr)
set.seed(1)

# --- data ---------------------------------------------------------------------
# Carlsson and Sinclair, Experiment 1 (Experiment1-w.dta, https://osf.io/qcdgp/)
file_cs <- tempfile(fileext = ".dta")
download.file("https://osf.io/download/wb4xy/", file_cs, mode = "wb")
dat  <- as.data.frame(read_dta(file_cs))
dat2 <- dat %>%
  dplyr::select(discW = job, beliefW = gwdisc, stigma, feminist = fem,
                beliefM = gmdisc, gender) %>%
  mutate(across(-gender, ~ as.numeric(scale(.))))
vars <- c("discW", "beliefW", "feminist", "stigma", "beliefM", "gender")
dat  <- dat2[complete.cases(dat2[, vars]), vars]
dat$gender <- as.factor(dat$gender)

# European Values Study, Germany (regression_EVS_Germany.csv, https://osf.io/7q5pf/)
EVS_Germany <- read.table("https://osf.io/download/j3zqh/",
                          header = TRUE, sep = ",")
EVS_Germany <- EVS_Germany[complete.cases(EVS_Germany), ]
EVS_Germany$attitude  <- c(scale(EVS_Germany$attitude))
EVS_Germany$education <- c(scale(EVS_Germany$education))
EVS_Germany$income    <- c(scale(EVS_Germany$income))
EVS_Germany$class     <- c(scale(EVS_Germany$class))
EVS_Germany$gender    <- as.factor(EVS_Germany$gender)

# --- the two applications --------------------------------------------------------
spec <- data.frame(label = c("CS", "EVS"),
                   data  = c("dat", "EVS_Germany"),
                   y     = c("discW", "attitude"),
                   x1    = c("feminist", "education"),
                   x2    = c("stigma", "class"))

# continuous predictors per data set (gender is a factor in both)
cont <- list(dat         = c("beliefW", "feminist", "stigma", "beliefM"),
             EVS_Germany = c("education", "income", "class"))

# prior probabilities of the equality structures uu, 0u, u0, 00
PM <- c(uu = 1/4, zu = 1/4, uz = 1/4, zz = 1/4)      # binomial, delta = 1/2
# PM <- c(uu = 1/3, zu = 1/6, uz = 1/6, zz = 1/3)    # uniform prior on delta

L   <- rbind(c(1, -1), c(0, 1))                      # (b1 - b2, b2) > 0
res <- NULL

for (a in 1:nrow(spec)) {

  d     <- get(spec$data[a])
  y     <- spec$y[a]
  x1    <- spec$x1[a]
  x2    <- spec$x2[a]
  preds <- c(cont[[spec$data[a]]], "gender")

  # --- BFpack: AFBF -------------------------------------------------------------
  fit <- lm(as.formula(paste(y, "~", paste(preds, collapse = " + "))), data = d)
  inp <- list()
  tab <- BF(fit, hypothesis = paste0(x1, " > ", x2, " > 0; ", x1, " = ", x2, " = 0"),
            BF.type = "AFBF")$BFtable_confirmatory
  Bzu <- BF(fit, hypothesis = paste0(x1, " = 0"), BF.type = "AFBF")$BFtable_confirmatory[1, "BF"]
  Buz <- BF(fit, hypothesis = paste0(x2, " = 0"), BF.type = "AFBF")$BFtable_confirmatory[1, "BF"]
  inp$AFBF <- c(c1  = unname(tab[1, "comp_O"]),   # relative complexity, order part
                f1  = unname(tab[1, "fit_O"]),    # relative fit, order part
                Bzu = unname(Bzu), Buz = unname(Buz),
                Bzz = unname(tab[2, "BF"]))       # 00 against uu

  # --- BayesFactor: JZS -----------------------------------------------------------
  Muu <- lmBF(as.formula(paste(y, "~", paste(preds, collapse = " + "))),
              data = d, progress = FALSE)
  Mzu <- lmBF(as.formula(paste(y, "~", paste(setdiff(preds, x1), collapse = " + "))),
              data = d, progress = FALSE)
  Muz <- lmBF(as.formula(paste(y, "~", paste(setdiff(preds, x2), collapse = " + "))),
              data = d, progress = FALSE)
  Mzz <- lmBF(as.formula(paste(y, "~", paste(setdiff(preds, c(x1, x2)), collapse = " + "))),
              data = d, progress = FALSE)
  buu  <- extractBF(Muu)$bf
  post <- posterior(Muu, iterations = 5e4, progress = FALSE)
  b1   <- post[, paste0(x1, "-", x1)]
  b2   <- post[, paste0(x2, "-", x2)]

  # relative complexity under the JZS prior: prior covariance of the continuous
  # slopes proportional to (Xc'Xc)^{-1}; the constraint is scale invariant
  Xc    <- scale(as.matrix(d[, cont[[spec$data[a]]]]), center = TRUE, scale = FALSE)
  V     <- solve(crossprod(Xc))[c(x1, x2), c(x1, x2)]
  c1jzs <- pmvnorm(lower = c(0, 0), upper = c(Inf, Inf), sigma = L %*% V %*% t(L))[1]

  inp$JZS <- c(c1 = c1jzs, f1 = mean(b1 > b2 & b2 > 0),
               Bzu = extractBF(Mzu)$bf / buu,
               Buz = extractBF(Muz)$bf / buu,
               Bzz = extractBF(Mzz)$bf / buu)

  # --- posterior probabilities ------------------------------------------------------
  for (m in names(inp)) {

    c1 <- inp[[m]]["c1"]; f1 <- inp[[m]]["f1"]
    B  <- c(1, inp[[m]]["Bzu"], inp[[m]]["Buz"], inp[[m]]["Bzz"])
    names(B) <- names(PM)

    # posterior probabilities of the equality structures
    PMy <- PM * B / sum(PM * B)

    # complexity-induced prior: H = {H1, H2, H3} and H* = {H1*, H2*}
    comp  <- c(PMy["uu"] * f1, PMy["zz"], 1 - PMy["uu"] * f1 - PMy["zz"])
    compS <- c(PMy["zz"], 1 - PMy["zz"])

    # uniform prior: complement under uu, normalized Bayes factors against uu
    BFu   <- c(f1 / c1, B["zz"], (1 - f1) / (1 - c1))
    unif  <- BFu / sum(BFu)
    unifS <- c(B["zz"], 1) / (B["zz"] + 1)

    res <- rbind(res,
                 data.frame(data = spec$label[a], hyp = paste(x1, ">", x2),
                            b1 = coef(fit)[x1], b2 = coef(fit)[x2],
                            method = m, prior = c("comp", "unif"), c1 = c1, f1 = f1,
                            H1 = c(comp[2], unif[2]), H2 = c(comp[1], unif[1]),
                            H3 = c(comp[3], unif[3]),
                            H1s = c(compS[1], unifS[1]), H2s = c(compS[2], unifS[2]),
                            Puu = PMy["uu"], Pzu = PMy["zu"], Puz = PMy["uz"], Pzz = PMy["zz"],
                            incoherent = c(NA, unname(unif[1] > unifS[2])),
                            row.names = NULL))
  }
}

num <- sapply(res, is.numeric)
out <- res
out[num] <- round(res[num], 3)
print(out, row.names = FALSE)
\end{verbatim}
}

\end{document}